\documentclass[
  journal=pasa,
  manuscript=research-paper, %% or "review"
  year=2020,
  volume=37,
]{cup-journal}

\usepackage{microtype,siunitx,booktabs}
\usepackage{caption}
\usepackage{ccaption}
\usepackage{subcaption}
\usepackage{verbatim}  % for commenting entire paras etc
\usepackage{float}
\usepackage{xcolor}     % to get change color of text for highlighting 
\usepackage{siunitx}
\usepackage{rotating}
\usepackage{amsfonts}
\usepackage{amssymb}
\usepackage[hidelinks]{hyperref}
\usepackage{cleveref}
\crefname{section}{§}{§§}
\Crefname{section}{§}{§§}
\usepackage{longtable}
\usepackage{natbib}
\usepackage{verbatim}
\usepackage{lscape}
\usepackage{pdflscape}
\usepackage{graphics}
\usepackage{epsfig}
\usepackage{multirow}
\usepackage{setspace}
\usepackage{bigdelim}
\usepackage{bigstrut}
\usepackage{floatflt}	%For floating figures/tables on one side
\usepackage{wrapfig}
\usepackage{multicol}

\usepackage{indentfirst}
\usepackage{tablefootnote}

\newcommand{\arcsec}{$''$}

\makeatletter
\newcommand*\mysize{%
   \@setfontsize\mysize{6.8}{10.0}%
}
\makeatother

\sffamily
\def\h1{H\,{\sc i}}

\def\c1{C\,{\sc i}}

\def\NH3{NH$_{3}$}
\def\ch3cn{CH$_{3}$CN}

\def\deg{$^{\circ}$}

\def\kms{km s$^{-1}$}

\def\apjs{ApJSS}
\def\apjl{ApJL}

\def\aap{A\&A}

\defcitealias{obreschkow14}{OG14}
\defcitealias{obreschkow16}{O16}
\defcitealias{Murugeshan2019}{M19}
\defcitealias{Murugeshan20}{M20}

\title[WALLABY Pilot Survey: PDR2]{WALLABY Pilot Survey: Public data release of $\sim 1800$ \h1 sources and high-resolution cut-outs from Pilot Survey Phase 2}

\author{C. Murugeshan}
\affiliation{ATNF, CSIRO, Space and Astronomy, PO Box 1130, Bentley, WA 6102, Australia}
\alsoaffiliation{ARC Centre of Excellence for All Sky Astrophysics in 3 Dimensions (ASTRO 3D), Australia}
\email[T. Westmeier]{tobias.westmeier@uwa.edu.au}

\author{N. Deg}
\affiliation{Department of Physics, Engineering Physics, and Astronomy, Queen’s University, Kingston, ON, K7L 3N6, Canada}

\author{T. Westmeier}
\affiliation{International Centre for Radio Astronomy Research (ICRAR), The University of Western Australia, 35 Stirling Highway, Crawley, WA 6009, Australia}
\alsoaffiliation{ARC Centre of Excellence for All Sky Astrophysics (ASTRO 3D), Australia}

\author{A. X. Shen}
\affiliation{ATNF, CSIRO, Space and Astronomy, PO Box 1130, Bentley, WA 6102, Australia}

\author{B. -Q. For}
\affiliation{International Centre for Radio Astronomy Research (ICRAR), The University of Western Australia, 35 Stirling Highway, Crawley, WA 6009, Australia}
\alsoaffiliation{ARC Centre of Excellence for All Sky Astrophysics (ASTRO 3D), Australia}

\author{K. Spekkens}
\affiliation{Department of Physics and Space Science, Royal Military College of Canada, PO Box 17000, Station Forces, Kingston, Ontario, Canada, K7K 7B4}
\alsoaffiliation{Department of Physics, Engineering Physics, and Astronomy, Queen’s University, Kingston, ON, K7L 3N6, Canada}

\author{O. I. Wong}
\affiliation{ATNF, CSIRO, Space and Astronomy, PO Box 1130, Bentley, WA 6102, Australia}
\alsoaffiliation{International Centre for Radio Astronomy Research (ICRAR), The University of Western Australia, 35 Stirling Highway, Crawley, WA 6009, Australia}
\alsoaffiliation{ARC Centre of Excellence for All Sky Astrophysics (ASTRO 3D), Australia}

\author{L. Staveley-Smith}
\affiliation{International Centre for Radio Astronomy Research (ICRAR), The University of Western Australia, 35 Stirling Highway, Crawley, WA 6009, Australia}
\alsoaffiliation{ARC Centre of Excellence for All Sky Astrophysics (ASTRO 3D), Australia}

\author{B. Catinella}
\affiliation{International Centre for Radio Astronomy Research (ICRAR), The University of Western Australia, 35 Stirling Highway, Crawley, WA 6009, Australia}
\alsoaffiliation{ARC Centre of Excellence for All Sky Astrophysics (ASTRO 3D), Australia}

\author{K. Lee-Waddell}
\affiliation{International Centre for Radio Astronomy Research (ICRAR), The University of Western Australia, 35 Stirling Highway, Crawley, WA 6009, Australia}
\alsoaffiliation{ATNF, CSIRO, Space and Astronomy, PO Box 1130, Bentley, WA 6102, Australia}
\alsoaffiliation{International Centre for Radio Astronomy Research (ICRAR), Curtin University, Bentley, WA 6102, Australia}

\author{H. D\'enes}
\affiliation{School of Physical Sciences and Nanotechnology, Yachay Tech University, Hacienda San José S/N, 100119, Urcuquí, Ecuador }

\author{J. Rhee}
\affiliation{International Centre for Radio Astronomy Research (ICRAR), The University of Western Australia, 35 Stirling Highway, Crawley, WA 6009, Australia}

\author{L. Cortese}
\affiliation{International Centre for Radio Astronomy Research (ICRAR), The University of Western Australia, 35 Stirling Highway, Crawley, WA 6009, Australia}
\alsoaffiliation{ARC Centre of Excellence for All Sky Astrophysics (ASTRO 3D), Australia}

\author{S. Goliath}
\affiliation{Canadian Astronomy Data Centre, NRC Herzberg, 5071 West Saanich Road, Victoria, British Columbia, Canada, V9E 2E7.}

\author{R. Halloran}
\affiliation{Queens University, 99 University Ave, Kingston, ON, K7L3N6, Canada }

\author{J. M. van der Hulst}
\affiliation{Kapteyn Astronomical Institute, P.O. Box 800, 9700AV Groningen, The Netherlands}

\author{P. Kamphuis}
\affiliation{Ruhr University Bochum, Faculty of Physics and Astronomy, Astronomical Institute (AIRUB), 44780 Bochum, Germany}

\author{B. S. Koribalski}
\affiliation{Australia Telescope National Facility, CSIRO, Space and Astronomy, P.O. Box 76, Epping, NSW 1710, Australia}
\alsoaffiliation{School of Science, Western Sydney University, Locked Bag 1797, Penrith, NSW 2751, Australia}

\author{R. C. Kraan-Korteweg}
\affiliation{Department of Astronomy, University of Cape Town, Private Bag X3, Rondebosch 7701, South Africa}

\author{F. Lelli}
\affiliation{INAF, Arcetri Astrophysical Observatory, Largo E. Fermi 5, Florence 50125, Italy}

\author{P. Venkataraman}
\affiliation{CIRADA, Dunlap Institute for Astronomy and Astrophysics, University of Toronto, Toronto, ON M5S 3H4, Canada}

\author{L. Verdes-Montenegro}
\affiliation{Instituto de Astrofísica de Andalucía (CSIC), Spain}

\author{N. Yu}
\affiliation{National Astronomical Observatories, Chinese Academy of Sciences, 20A Datun Rd, Chaoyang District, Beijing 100101, China}
\alsoaffiliation{Key Laboratory of Radio Astronomy and Technology, Chinese Academy of Sciences, 20A Datun Rd, Chaoyang District, Beijing 100101, China}

\doi{10.1017/pasa.2020.32}

\received {dd Mmm YYYY}
\revised  {dd Mmm YYYY}
\accepted {dd Mmm YYYY}
\published{22 September 2020}

 %% First letter not capped

\begin{document}
%\input{example-content}

% Abstract of the paper
\begin{abstract}
We present the Pilot Survey Phase 2 data release for the Wide-field ASKAP L-band Legacy
All-sky Blind surveY (WALLABY), carried-out using the Australian SKA Pathfinder (ASKAP). We present 1760 \h1 detections (with a default spatial resolution of 30\arcsec) from three pilot fields including the NGC 5044 and NGC 4808 groups as well as the Vela field, covering a total of $\sim 180$ deg$^2$ of the sky and spanning a redshift up to $z \simeq 0.09$. This release also includes kinematic models for over 126 spatially resolved galaxies. The observed median rms noise in the image cubes is 1.7 mJy per 30\arcsec\ beam and 18.5 kHz channel. This corresponds to a 5$\sigma$ \h1 column density sensitivity of $\sim 9.1\times10^{19}(1 + z)^4$ cm$^{-2}$ per 30\arcsec\ beam and $\sim 20$ \kms channel, and a 5$\sigma$ \h1 mass sensitivity of $\sim 5.5\times10^8 (D/100$ Mpc)$^{2}$ M$_{\odot}$ for point sources. Furthermore, we also present for the first time 12\arcsec\ high-resolution images (``cut-outs") and catalogues for a sub-sample of 80 sources from the Pilot Survey Phase 2 fields. While we are able to recover sources with lower signal-to-noise ratio compared to sources in the Public Data Release 1, we do note that some data quality issues still persist, notably, flux discrepancies that are linked to the impact of side lobes associated with the dirty beams due to inadequate deconvolution. However, in spite of these limitations, the WALLABY Pilot Survey Phase 2 has already produced roughly a third of the number of HIPASS sources, making this the largest spatially resolved \h1 sample from a single survey to date. 

\end{abstract}

%%%%%%%%%%%%%%%%% BODY OF PAPER %%%%%%%%%%%%%%%%%%

\section{Introduction}

The role of neutral hydrogen (\h1) gas as the primary fuel for star formation in galaxies is now well-established. Several surveys utilizing single-dish (e.g.,~\citealt{Meyer2004};~\citealt{Koribalski2004};~\citealt{Wong2006};~\citealt{Giovanelli2005};~\citealt{Catinella2010}) and interferometric observations (e.g.,~\citealt{vanderhulst2001};~\citealt{Verdes-Montenegro2005};~\citealt{Walter2008};~\citealt{Begum2008};~\citealt{Heald2011};~\citealt{Cappellari2011};~\citealt{Hunter2012};~\citealt{Ott2012};~\citealt{Koribalski2018}), have
shown the significance of the \h1 gas in understanding galaxy evolution. While significant progress has been made in studying galaxy evolution through resolved \h1 observations, a thorough perspective of the \h1 gas distribution in galaxies, its statistical properties and its relation to star formation necessitates more resolved observations of tens of thousands of galaxies from unbiased surveys. %New and upcoming \h1 surveys using advanced radio interferometers like ASKAP \citep{Hotan21}, MeerKAT \citep{Jonas2016}, upgraded VLA \citep{Perley2011}, and GMRT \citep{Gupta2017} will start to address this need until the SKA-mid project is underway.

The Wide-field ASKAP L-band Legacy All-sky Blind surveY (WALLABY;~\citealt{Koribalski2020}) is already contributing on this front and is expected to detect over $\sim 200,000$ sources out to a redshift of $z \sim 0.1$ covering the majority of the southern sky using the Australian SKA Pathfinder (ASKAP;~\citealt{Hotan21}) telescope. This is almost a factor of 10 better than the number of sources detected in ALFALFA (\citealt{Giovanelli2005};~\citealt{Haynes2018}). In addition, WALLABY will be able to resolve tens of thousands of galaxies with a default resolution of 30\arcsec, while also producing higher-resolution 12\arcsec\ `cut-outs' for a select sub-sample of galaxies. 
The 12\arcsec\ data products will become part of regular full WALLABY survey data releases. The aim is to image all HIPASS sources ($N \sim 5000$) in high resolution, in addition, the WALLABY team is compiling a catalogue of galaxies selected based on their optical properties which we also intend to image at 12\arcsec\ resolution. As such, WALLABY will deliver 12\arcsec\ data products for thousands of galaxies in its first 5-year survey period. Some of the main goals that can be achieved with the higher resolution data include but are not limited to:

\begin{itemize}
    \item Studying the \h1 morphology of galaxies at higher resolution and detailed kinematic studies of local galaxies by accurately modelling the \h1 distribution. In addition, the higher resolution also allows for complementary studies with IFU observations. This will also enable us to look for kinematical misalignment between the \h1 gas and/or the ionised gas and stars in galaxies (e.g.,~\citealt{Wong2015};~\citealt{Bryant2019}).
    \item Dynamical scaling laws of rotation supported galaxies and resolved angular momentum studies (e.g.,~\citealt{McGaugh2000};~\citealt{Verheijen2001};~\citealt{Lelli2017};~\citealt{Murugeshan2020};~\citealt{Kurapati2018};~\citealt{Mancera_Pina2021};~\citealt{Sorgho2024}) and tracing the effects of non-axisymmetric potentials such as bars and bulges on the \h1 gas in galaxies (e.g.,~\citealt{Masters2012};~\citealt{Murugeshan2023}). With the higher resolution, we may also be able to trace warps in the discs of galaxies more accurately \citep{2002Garcia_Ruiz}.
    \item Probing the dynamics of galaxies using reliable and robust rotation curves derived from the higher resolution data (\citealt{de_Blok2008};~\citealt{Lelli2012}). This will also enable us to probe the dark matter distribution in local galaxies and additionally address the core-cusp problem relating to dwarf galaxies \citep{Katz2017}.
    \item Studying the star formation properties and star formation laws pertaining to the high column density \h1 gas (N$_{\textrm{\h1}} \geq 10^{20}$ cm$^{-2}$). In addition, we may also be able to probe the \h1 gas and star formation properties of well resolved local dwarf galaxies (e.g.,~\citealt{Roychowdhury2014};~\citealt{Bacchini2020}).
\end{itemize}

These science cases highlight the need for high-resolution \h1 imaging of targeted (and potentially interesting) galaxies. As such, WALLABY will truly pave the way for high-resolution \h1 studies of local galaxies to an unprecedented scale by imaging thousands of galaxies at 12\arcsec\ resolution.

For more specific details on the WALLABY survey we refer the reader to the original WALLABY paper \citep{Koribalski2020}. We summarise some important updated WALLABY survey parameters for the next 5-year period in Table~\ref{tab:WALLABY_params}. Pre-pilot and pilot surveys were conducted to assess ASKAP data quality and to plan full survey strategies. The targeted fields of the pre-pilot surveys are listed in the following WALLABY pre-pilot survey papers by ~\citet{For2021},~\citet{Wong2021} and ~\citet{Murugeshan2021}, while the details of the public data release of the Pilot Survey Phase 1 (hereafter Phase 1 or PDR1) observations are described in \citet{westmeier2022} and \citet{Deg2022}. 

\begin{table}
\caption{Important updated WALLABY survey parameters}
\begin{tabular}{ll}
\hline\hline
Parameter & Value \\
\hline
Sky coverage & 14000 deg$^2$ \\
Declination range (core survey) & -60\deg $\leq \delta \leq$ -15\deg  \\
No. of ASKAP tiles & 552 \\
Integration time per tile (h) & 16 \\
Maximum baseline -- default [high-resolution] (km) & 2 [6] \\
Spatial resolution (FWHM) -- default [high-resolution] & 30\arcsec\ [12\arcsec] \\
Redshift range & 0 -- 0.1 \\
Observed Bandwidth (MHz) & 1151.5 -- 1439.5 \\
Frequency resolution (kHz) &	18.5 \\
Velocity resolution (at $z = 0$) (\kms) & 4 \\
RMS flux density sensitivity (per 0.1 MHz/20 \kms)	& 0.7 mJy \\
RMS column density sensitivity (per 0.1 MHz/20 \kms) &	$1.7 \times 10^{19}$ cm$^{-2}$ \\
\hline
\end{tabular}
\label{tab:WALLABY_params}
\end{table}

In this paper we present the public data release of the \h1 catalogues and associated data products from the WALLABY Pilot Survey Phase 2 (hereafter also Phase 2 or PDR2) observations. Section~\ref{sec:Data&Methods} gives details of the targeted fields, observations, data reduction and briefly introduces the methods employed for the validation of the observations. In Section~\ref{sec:Source_finding} we highlight the source finding strategy and provide specific notes for each target field. In Section~\ref{sec:source_characterization}, we present the general properties of the detected 30\arcsec\ sample. Section~\ref{sec:postage_stamps} introduces the high-resolution 12\arcsec\ data, the data reduction pipeline and characteristics of the sources. In Section~\ref{sec:flux_discrepancy} we describe an observed flux discrepancy in the WALLABY data and give details on the simulation studies undertaken to uncover the origins of this flux discrepancy. Section~\ref{sec:3D_modelling} describes the kinematic modelling pipeline and presents the kinematic models along with some comparisons between the 30\arcsec\ and 12\arcsec\ models. Finally, Section~\ref{sec:data_access} provides details on how to access the data, while in Section~\ref{sec:Summary_Future} we provide a summary and the future goals of the WALLABY survey. 

\section{Observations and data reduction}
\label{sec:Data&Methods}

The data used in this work has been acquired via ASKAP observations of the WALLABY Pilot Survey Phase 2 fields -- NGC 4808, NGC 5044 and the Vela group. Located at the Inyarrimanha Ilgari Bundara, the Murchison Radio-astronomy Observatory (MRO), ASKAP \citep{Hotan21} is a state-of-the-art radio interferometer comprising of 36 12-meter antennas, equipped with Mk~II phased array feeds (PAFs; \citealt{DeBoer09};~\citealt{Chippendale10};~\citealt{hotan14}). ASKAP is able to form 36 beams simultaneously on the sky using the advantage of the PAF, thus covering a very large area on the sky in a single pointing. For WALLABY, the 36 beams are typically arranged in the form of $6 \times 6$ square footprints (see Figure~\ref{fig:field_selection1}). The simultaneous field of view (FOV) of ASKAP is $\sim 30$ deg$^{2}$ at 1.4 GHz. For the WALLABY survey observations, two $6 \times 6$ square footprints (footprint A and B) are interleaved to attain the required uniform sensitivity across the field. A combination of both footprints A and B is referred to as a tile.

The observations of the various Phase 2 fields were carried out between April 2021 and May 2022 (for exact observing dates refer to Table~\ref{tab:Obervations}) with an integration time of $\sim 8$ h for each footprint and thus a total on-source time of $\sim 16$ h per tile. During the observations, most of the 36 antennas were used to correlate the data, although a few antennas were flagged as bad during the data reduction process (for details, refer to Table.~\ref{tab:Obervations}). 

We note that the observations were carried out in the frequency range of 1152 -- 1440 MHz, with a total bandwidth of 288 MHz, consisting of 15,552 channels corresponding to a spectral resolution of 18.5 kHz. As with Phase 1 observations, we note that only the upper half of the band above $\sim 1300$ MHz has been processed as the observations below this frequency are severely affected by Radio Frequency Interference (RFI) due to  Global Positioning System/Global Navigation Satellite System (GPS/GNSS).

\subsection{Field selection}
\label{sec:field_selection}

\begin{figure*}
    %\centering
    \vspace*{-2.0cm}
    \includegraphics[width=1\columnwidth]{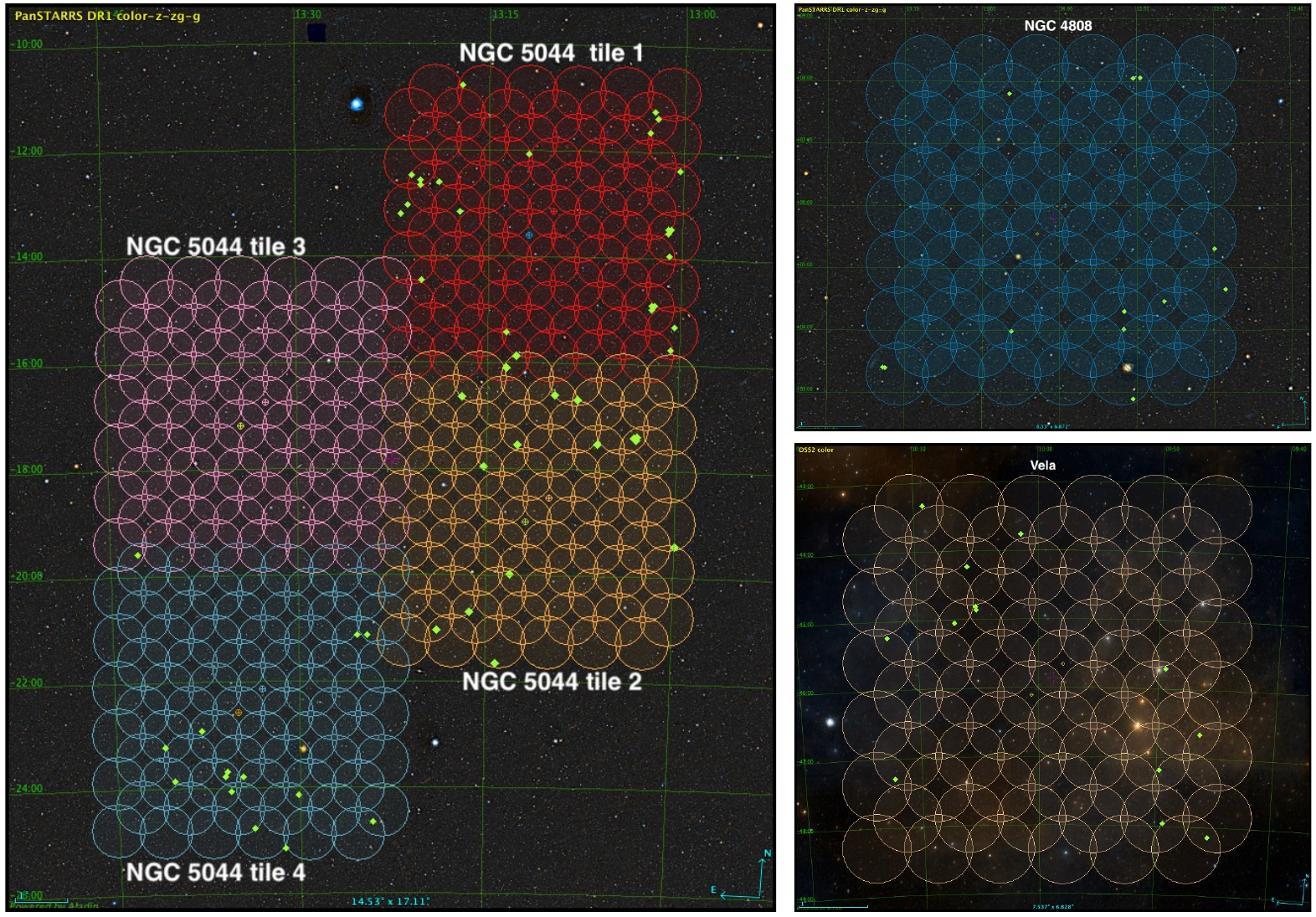}     
    \caption{The ASKAP footprints covering the Pilot Phase 2 fields overlaid on top of their PanSTARRS composite optical images. The green points show the location of the HIPASS sources imaged with a 12\arcsec\ resolution for the high-resolution cut-outs.}
    \label{fig:field_selection1}
\end{figure*}

For the Phase 2 observations, each ASKAP Science Survey Project (SSP) was allocated a total of 100 h of observing time. In Figure~\ref{fig:field_selection1}, we show the targeted pilot Phase 2 fields. The field selection was decided based on the following criteria: \\

\noindent \textbf{Scientific merit} -- The Phase 2 fields were chosen on their merit, ensuring that multi-wavelength data is readily available, and in addition, have the potential to maximise the science goals, which include probing large scale structures in the zone of avoidance (ZOA) and investigating environmental effects on galaxy groups. \\ 

\noindent \textbf{Commensality with other ASKAP Science Survey Teams} -- WALLABY is commensal with other ASKAP surveys such as the Evolutionary Map of the Universe (EMU;~\citealt{Norris2011}) survey, Polarisation Sky Survey of the Universe's Magnetism (POSSUM;~\citealt{Gaensler2010}), the Galactic ASKAP Survey (GASKAP;~\citealt{Dickey2013}) and the Commensal Real-time ASKAP Fast Transients Survey (CRAFT;~\citealt{Macquart10}). The NGC 5044 tile 3 field was chosen to be the EMU-POSSUM-WALLABY three-way commensal field. While the Vela field was chosen to be commensal with GASKAP, wherein observations in the Galactic range ($V_{sys} < 500$ \kms) were reduced in `zoom mode' with a full spectral resolution of 2 \kms. \\
    
\noindent \textbf{Source finding strategy} -- The NGC 5044 fields were targeted as they cover a contiguous region on the sky (see Figure~\ref{fig:field_selection1}). Observing overlapping fields/tiles was necessary so as to test our source finding strategy in preparation for the full survey, which will involve running the source finding pipeline on contiguous adjacent fields. Refer to Section~\ref{subsec:source_find_strategy} for more details on the source finding strategy. 

\begin{table*}
\caption{Details of the observations. Col (1): Name of the field; Col (2): tile/footprint; Col (3): ASKAP Scheduling block identifier (SBID) used to tag the data in CASDA; Col (4): Date of observation; Col (5) - (6): RA and Dec of the centre of the footprint, respectively, in J2000; Col (7): Phase rotation of the footprint on the sky in deg; Col (8): Number of antennas used; Col (9): Flagged fraction. $^a$EMU-POSSUM-WALLABY commensal field; $^b$GASKAP-WALLABY commensal field.}
\begin{tabular}{lllllllll}
\hline\hline
% this row describes the columns
Field  & Footprint & SBID  &  Date  & RA & Dec  & $\phi$ & No. of & Flagged \\
     &  &   &  & (J2000)  & (J2000) & (deg) & Ant. & fraction \\
(1)  & (2)  & (3) &  (4)  & (5) & (6)  & (7) & (8) & (9) \\
\hline\hline
NGC 5044 & 1A & 33879 & 2021 Nov 23 & $13^{\rm h}10^{\rm m}00^{\rm s}$ & $-13^{\circ}09'04''$ & 0.00 & 34 & $\sim 10$\% \\
         & 1B & 34302 & 2021 Dec 9 & $13^{\rm h}11^{\rm m}51^{\rm s}$ & $-13^{\circ}36'02''$ & -0.11  & 33 & $\sim 20$\% \\
         & 2A & 34166 & 2021 Dec 4 & $13^{\rm h}10^{\rm m}00^{\rm s}$ & $-18^{\circ}32'53''$ & 0.00  & 32 & $\sim 20$\% \\
         & 2B & 34275 & 2021 Dec 8 & $13^{\rm h}11^{\rm m}54^{\rm s}$ & $-18^{\circ}59'50''$ & -0.15  & 31 & $\sim 20 - 25$\% \\
         & 3A$^a$ & 31536 & 2021 Sep 1 & $13^{\rm h}32^{\rm m}30^{\rm s}$ & $-16^{\circ}45'00''$ & -1.20  & 33 & $\sim 10$\% \\
         & 3B$^a$ & 40905 & 2022 May 23 & $13^{\rm h}34^{\rm m}25^{\rm s}$ & $-17^{\circ}11'24''$ & -1.34  & 33 & $\sim 10$\% \\
         & 4A & 25701 & 2021 Apr 4 & $13^{\rm h}33^{\rm m}00^{\rm s}$ & $-22^{\circ}08'48''$ & -1.20  & 30 & $\sim 25 - 30$\% \\
         & 4B & 25750 & 2021 Apr 5& $13^{\rm h}34^{\rm m}59^{\rm s}$ & $-22^{\circ}35'12''$ & -1.39  & 30 & $\sim 25 - 30$\% \\
\hline
NGC 4808 & A & 33681 & 2021 Nov 18 & $12^{\rm h}59^{\rm m}37^{\rm s}$ & $+06^{\circ}00'00''$ & 0.00 & 34 & $\sim 5 - 10$\% \\
         & B & 37604 & 2022 Feb 23& $13^{\rm h}01^{\rm m}26^{\rm s}$ & $+05^{\circ}32'59''$ & 0.05  & 34 & $\sim 5 - 10$\% \\
\hline 
Vela$^b$ & A & 33521 & 2021 Nov 14 & $09^{\rm h}58^{\rm m}00^{\rm s}$ & $-45^{\circ}39'17''$ & 0.00  & 34 & $\sim 10$\% \\
         & B & 33596 & 2021 Nov 16& $10^{\rm h}00^{\rm m}36^{\rm s}$ & $-46^{\circ}06'10''$ & -0.47  & 34 & $\sim 10$\% \\
\hline
%Virgo South & A & 2 Apr 2021 & $12^{\rm h}57^{\rm m}04^{\rm s}$ & $-05^{\circ}$59'19'' & 0.00 & 8.0 & 25612 & Bright continuum quality gate field \\
%            & B & 3 Apr 2021 & $12^{\rm h}58^{\rm m}52.682^{\rm s}$ & $-06^{\circ}$26'18.28'' & -0.05 & 8.0 & 25654 & Failed to pass validation stage as both \\
% & & & & & & & & footprints A \& B plagued by continuum artefacts \\
\hline
\end{tabular}
\label{tab:Obervations}
\end{table*}

\subsection{Default 30\arcsec\ WALLABY data reduction pipeline}
\label{subsec:data_reduction}

For a detailed description of the data reduction process of the default 30\arcsec\ WALLABY observations, we refer the reader to \citet{westmeier2022}. We describe very briefly the different stages of the default pipeline. We note that each of the steps below are performed independently for each ASKAP beam before they are mosaicked to form the final image cubes. First, the pipeline runs an automated flagging procedure which identifies bad antennas and flags the bad data for each beam. After the flagging procedure the pipeline proceeds to perform the bandpass calibration, followed by imaging the continuum. Then using the component and sky models derived from the continuum imaging, continuum subtraction is performed in the UV-domain. The next steps involve imaging each ASKAP beam separately, which also includes the deconvolution step, where the data is cleaned to a peak residual flux density of 3.5 mJy, followed by a deeper cleaning (within the pixels corresponding to the identified clean components) to a residual peak flux density threshold of 0.5 mJy. This is then followed by restoring the clean components convolved with a 30\arcsec\ Gaussian beam and adding back the residuals to form the image cubes. After the restoring phase, an image-based continuum subtraction routine is performed. A primary beam correction is then performed after which all the beams are mosaicked together to form two footprint (A and B) image cubes, which are then mosaicked to form the final full sensitivity image cube. We note that the main change in the data reduction pipeline for Phase 2 is the use of holographic measurements of the actual ASKAP primary beams used for the observations \citep{hotan16} for the primary beam correction, as opposed to the static Gaussian primary beam correction that was used for the Phase 1 data reduction. The introduction of the holography models for the correction provides more accurate primary beam model weights leading to more accurate flux recovery from detections across the entire FOV compared to the flux based on the static Gaussian primary beam model. 

\subsection{Data quality assessment and validation}
\label{sec:data_quality}

RFI and antenna flagging are performed on a beam-by-beam basis. The overall flagged visibility fraction ranges from 5 to 30\% across all beams, and typically all 36 antennas were utilised for all beams.

We evaluate the data quality of each footprint image cube based on a set of metrics. These metrics were established based on the data in the WALLABY early science field of M83 and pre-pilot field of Eridanus (see~\citealt{For2019};~\citealt{For2021}), which include RMS, minimum and maximum flux densities, 1 percentile noise level and median absolute deviation of median flux (MADMF). Each set consists of values for three types of image cubes, i.e., before and after continuum subtraction image cubes as well as a residual cube. 

The broadband RFI/artefacts are evaluated with the MADMF statistic. This metric is sensitive to strong artefacts. The distribution of flux density values for all voxels in each beam at the 1 percentile level indicates any bandpass calibration and/or sidelobe issues. All these metrics and observation information are presented in a HTML style summary report for each footprint. The report of each footprint and description of each metric is available at the CSIRO ASKAP Science Data Archive (CASDA;~\citealt{Huynh2020}).

Following this, a quality checking pipeline verifies that the footprints in CASDA meet the data quality requirements as mentioned above. The pipeline is executed when a new observation is available on CASDA. We run the Source Finding Application (SoFiA;~\citealt{Serra15};~\citealt{Westmeier21}) described in detail in Section~\ref{sec:Source_finding} on the mosaicked image cubes to generate moment 0 images of the field. Then, we verify by eye that there are no significant artefacts in the source finding output. Footprints that show significant artefacts from the source finding run are rejected by the team, and marked to be re-observed. Accepted footprints are recorded in a database (for more details see~\ref{appendix:manual_inspection_workflow}), which is used by the main source finding pipeline.

\section{Source Finding and parametrisation}
\label{sec:Source_finding} 

Source finding on the final image cubes was performed using the Source Finding Application (SoFiA;~\citealt{Serra15};~\citealt{Westmeier21}) version 2.3.1. For this purpose, each tile was split into sub-regions of approximately $1500 \times 1500$ spatial pixels and 1400 spectral channels for parallel processing on multiple nodes of the Nimbus computing cluster at the Pawsey Supercomputing Centre in Perth. In total, the frequency range of 1305 -- 1418~MHz, corresponding to a recession velocity range of $500 \lesssim \mathrm{c} z \lesssim 26,500~\mathrm{km \, s}^{-1}$, was searched for \h1~emission.

Each sub-region was first multiplied by the square root of the associated weights cube to normalise the noise across the data cube. This was followed by automatic flagging of artefacts and the positions of radio continuum sources with flux densities $> 150~\mathrm{mJy}$ in the Rapid ASKAP Continuum Survey (RACS;~\citealt{McConnell2020}) catalogue. A circle with a radius of 5 pixels (or 30\arcsec) was flagged around the position of each such continuum source, flagging the entire frequency range (including any \h1 emission) within those pixels, creating a circular hole in the affected area. If an \h1 detection is affected by flagging, the flag parameter in the catalogue is set accordingly to alert users of the fact that the detection is adjacent to flagged pixels. Additional noise normalisation in a running window of size $51 \times 51$ spatial pixels and 51 spectral channels was carried out to normalise any remaining noise variation that was not accurately reflected by the weights cube. In addition, a robust first-order polynomial was fitted to each spectrum to remove any remaining low-level continuum residuals, and SoFiA's ripple filter was employed to remove any low-level bandpass ripples due to RFI.

\begin{figure}
    \vspace*{-2cm}
    %\hspace*{-0.7cm}
    \includegraphics[width=1\columnwidth]{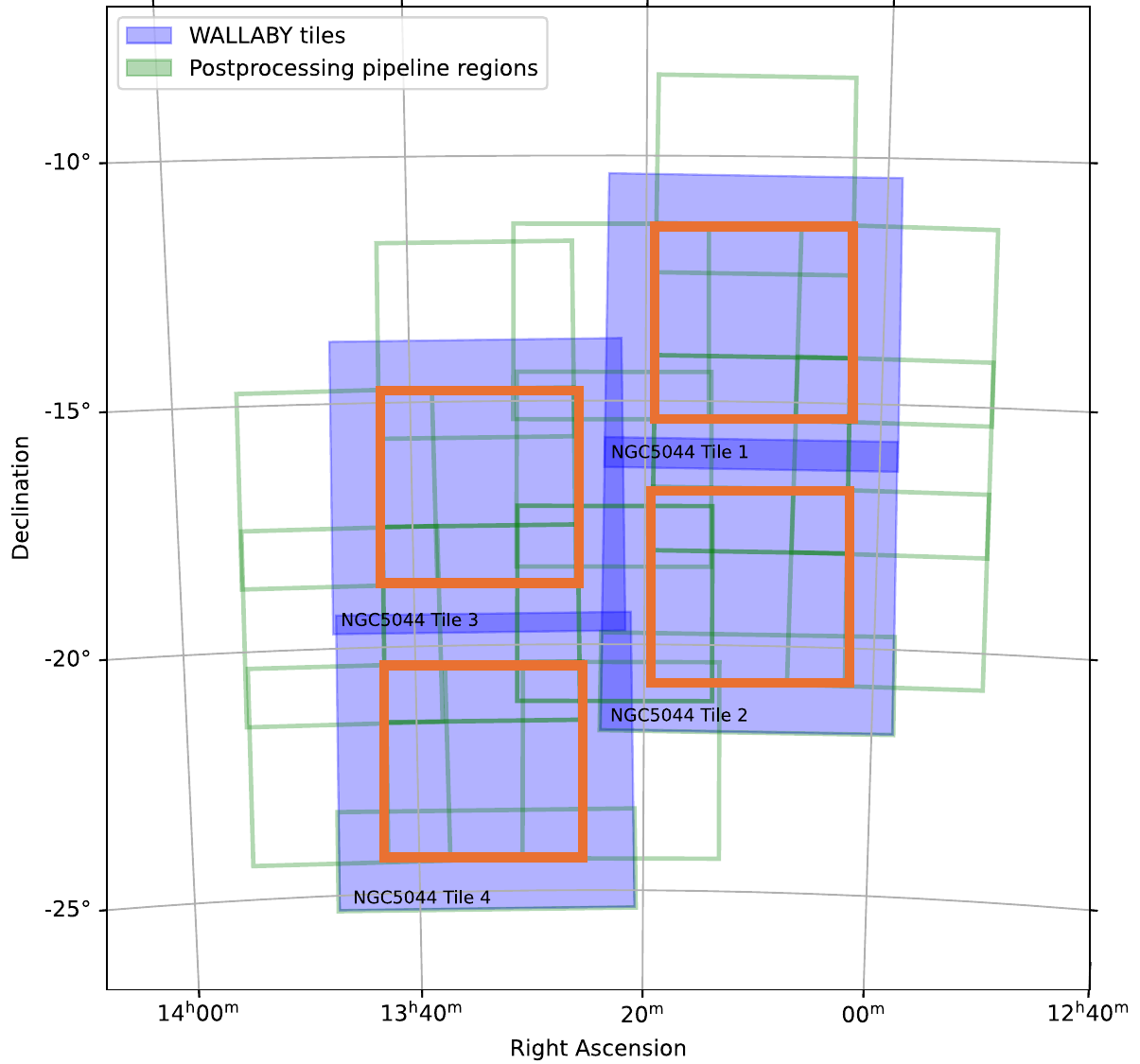}   
    \caption{Strategy for source finding in the NGC 5044 field which has overlapping regions. Tiles are shown as blue shaded regions while each orange box corresponds to a central $\sim 4^{\circ} \times 4^{\circ}$ area, where the source finding is performed. For the NGC 5044 field central regions are processed when both footprints have been observed, and overlapping regions are processed when adjacent tiles are completed. The light green boxes represent $\sim 4^{\circ} \times 4^{\circ}$ areas where source finding is run when appropriate adjacent tiles are available (or become available in the future).}
    \label{fig:source_finding_strategy}
\end{figure}

After these preconditioning steps, SoFiA's `smooth and clip' (S+C) algorithm was used to detect emission above a threshold of $3.8$ times the noise level in each smoothing iteration. Smoothing kernels of 0, 5 and 10 spatial pixels and 0, 3, 7, 15 and 31 spectral channels were employed to boost the signal-to-noise ratio of faint, extended \h1 emission on spatial scales of up to 1~arcmin and velocity widths of up to about $120~\mathrm{km \, s}^{-1}$ (at $z \approx 0$). All detected pixels were then merged into coherent detections across a merging length of 2 spatial pixels and 3 spectral channels, with detections smaller than 5 pixels or channels discarded. Next, SoFiA's reliability module was used to discard all detections with a reliability of less than 0.7 or an integrated signal-to-noise ratio of less than 3. In addition, detections with a total of less than 300 spatial and spectral pixels were also discarded as unreliable.

The remaining detections were then parameterised before SoFiA generated the final source catalogue and output products, including cubelets, moment maps and integrated spectra for all detections. Table~\ref{tab:SoFiA_params_30and12} in \ref{appendix:SoFiA_params} lists some important SoFiA parameter values used for the 30\arcsec\ source finding runs.

\subsection{Source finding strategy}
\label{subsec:source_find_strategy}

A pipeline has been developed through the Australian SKA Regional Centre (AusSRC) to run the source finding for the WALLABY survey. The pipeline communicates with external databases such as CASDA and the WALLABY database to automatically check for new footprints (and tiles) that have been uploaded on to CASDA. When a new observing tile has been deposited in CASDA, the pipeline mosaics overlapping regions of adjacent tiles outside the central $4^{\circ} \times 4^{\circ}$ (orange boxes in Figure~\ref{fig:source_finding_strategy}) and executes the source finding application.

The Vela and NGC~4808 fields were covered by only a single ASKAP tile each, and hence source finding was performed on the full tile at once. For the four-tile NGC~5044 mosaic we instead employed a staged source finding approach to account for the gradual completion and release of observations for this field. Figure~\ref{fig:source_finding_strategy} gives a visual representation of the source finding strategy adopted for the NGC~5044 field. First, SoFiA was run on the central $4^{\circ} \times 4^{\circ}$ region (orange boxes) of each individual NGC~5044 tile. This central region corresponds to the area across which the noise level is roughly constant in an individual tile. Beyond the central $4^{\circ} \times 4^{\circ}$ region, the noise in the outskirts of the tile typically tend to increase by a factor of two or more (see Appendix A in \citealt{westmeier2022}). Once adjacent tiles became available, SoFiA was then additionally run on the overlapping regions between those tiles in steps of $4^{\circ} \times 4^{\circ}$ regions (green boxes), to gradually build up a source catalogue of the entire NGC~5044 field. This staged source finding approach will also be applied to the full WALLABY survey in the future. The NGC~5044 mosaic provided us with the opportunity to develop and test this approach in anticipation of the full survey observations.

Detections from the source finding pipeline are uploaded into a database and WALLABY sources are then manually accepted following visual inspection. For more details on the manual workflow refer to \ref{appendix:manual_inspection_workflow}.

\subsection{Notes on individual fields}
\label{subsec:notes_on_fields}

In this section we present some pertinent notes on the individual fields. We note that while due care has been taken to avoid artefacts and false positives in the final source catalogues through visual inspection of all raw detections from SoFiA, we caution that some false positives may still remain in the final source catalogue as well as the issue of blended sources and/or sources broken into separate detections. This is true for all three Phase 2 fields. Where possible, comments are made in the source catalogue highlighting such issues. In addition, we also added `multiplet' and `component' tags to mark such cases.

\subsubsection{NGC 5044}
\label{subsubsec:Notes_NGC5044}

The data quality for the NGC 5044 tiles is good overall with only a few artefacts still present in the final mosaicked image cube.
This release consists of the source finding detections from the four NGC 5044 tiles covering $ 4 \times 30 \sim 120$ deg$^{2}$ and spanning a velocity range of $cz \sim 500 - 26,500$ \kms ($z < 0.089$) using the full RFI-free higher frequency band available for WALLABY. Some artefacts still remain in the data cube particularly related to faint continuum residuals and sidelobes that have affected the northern edge of tile 1, the southern edge of tile 2 and a small region of tile 4 of the NGC 5044 mosaic. We note that this may have reduced the completeness of the source finding runs in the affected regions.

After the source finding run, all detections were visually inspected and obvious artefacts were removed following which 1326 detections remain. We note that the NGC 5044 tile 4 was the only Phase 2 tile for which a Gaussian primary beam model was used for primary beam correction instead of using a holography model, due to which we anticipate minor flux-related issues such a potential increase in flux by about $\sim 15 - 20$\% for sources that lie further away from the beam centre and/or close to the edge of the tile/footprint. For tiles 1, 2 and 3 of the NGC 5044 field, the holography-based primary beam correction was performed. 

\subsubsection{NGC 4808}
\label{subsubsec:Notes_NGC4808}

The data release for the NGC 4808 field covers 30 deg$^{2}$ of the sky with a velocity range of $cz \sim 500 - 26,500$ \kms ($z < 0.089$). There were no major issues identified with the NGC 4808 field and the data quality is overall good, with very few artefacts in the image cube. The source finding run resulted in the retention of 231 detections following removal of few faint artefacts. 

\subsubsection{Vela}
\label{subsubsec:Notes_Vela}

The Vela field covers 30 deg$^{2}$ with a redshift range of $cz \sim 500 - 25,400$ \kms ($z < 0.085$). As mentioned in Section~\ref{sec:field_selection}, this field was observed commensally with the GASKAP-\h1 project in spectral zoom mode and processed at the full spectral channel width of 9.26 kHz. After this, the extragalactic frequency range of the data was re-binned to the default WALLABY spectral resolution of 18.5 kHz prior to spectral imaging. However, due to flagging preceding binning, some faint RFI from global navigational satellite systems was not fully flagged in the higher spectral resolution data, which has resulted in a significant number of false detections at frequencies of $\nu \simeq 1380$ MHz and $\nu \simeq 1310$ MHz. This therefore has resulted in the reliability of detections at those frequencies to be reduced which may have resulted in some genuine \h1 sources being omitted by SoFiA. Overall, 203 detections are retained after visual inspection and removing artefacts and false positives. 

%%%%%%%%%%%%%%%%%%%%%%%%%%%%%%%%%%%%%% Subsection: Source characterization    %%%%%%%%%%%%%%%%%%%%%%%%%%%%%%%%%%%%%%%%
\section{Source characterization}
\label{sec:source_characterization}

In this section we highlight some characteristics of the source properties from the Phase 2 source finding runs, such as the distribution of the signal-to-noise-ratio (SNR) of the detected sources, size distribution, \h1 mass distribution as well as their \h1 mass -- distance plot. We also compare the Phase 2 source properties with the Phase 1 detections in order to highlight the significant improvement in the data quality. 

\begin{figure*}
    %\centering
    \vspace*{-1.5cm}
    \includegraphics[width=1\textwidth]{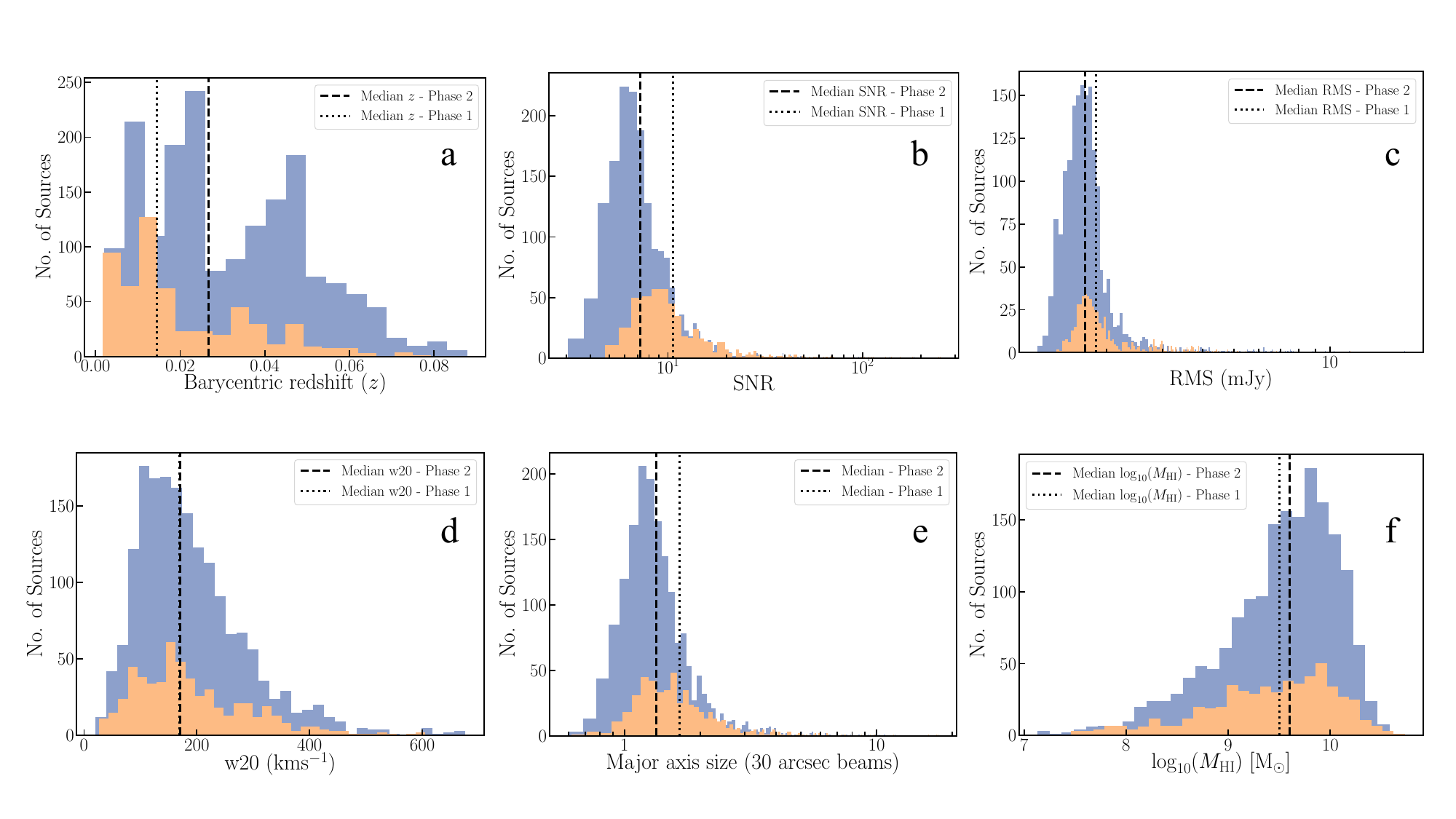}     
    \caption{\textbf{a)} Distribution of the barycentric redshifts of the Phase 2 sources (blue) compared to the Phase 1 detections (orange). \textbf{b)} Histogram of the Signal-to-noise (SNR) for both the Phase 2 and Phase 1 detections. \textbf{c)} Local noise distribution in the images cubes for the Phase 2 and Phase 1 detections. \textbf{d)} Distribution of the $w_{20}$ \h1 line-width distribution. \textbf{e)} Histogram of the major axis size (in units of 30\arcsec\ beams) for the two samples. \textbf{f)} The \h1 mass distribution for the Phase 2 and Phase 1 samples. In all plots, the dashed and dotted black lines represents the median value of the distribution for the Phase 2 and Phase 1 detections, respectively. }
    \label{fig:PDR2_properties}
\end{figure*}

Panel \textbf{a)} in Figure~\ref{fig:PDR2_properties} shows the distribution of the barycentric redshift for the Phase 2 detections (in blue) compared to the redshift distribution of sources in Phase 1. We find that the median redshift of the sources in Phase 2 is $\sim 0.027 ~(cz \sim 8094$~\kms). The median redshift of sources in the NGC 5044 field is $\sim 0.025 ~(cz \sim 7495$~\kms), the NGC 4808 field is $\sim 0.039 ~(cz \sim 11692$~\kms) and the Vela field is $\sim 0.04 ~(cz \sim 11992$~\kms). We see the clumping in redshifts in two distinct peaks in Figure~\ref{fig:PDR2_properties}. In comparison, the Phase 1 sources were mainly from nearby groups and clusters and as such, show a median barycentric redshift of $\sim 0.014 ~(cz \sim 4197$~\kms).

In panel \textbf{b)} of Figure~\ref{fig:PDR2_properties} we show the SNR (defined as the ratio of the integrated flux to the uncertainty in the integrated flux measured by SoFiA) of the detected sources for both the Phase 1 and Phase 2 fields. As reported in \citet{westmeier2022}, the peak of the SNR for the Phase 1 data is $\sim 9$ (with median $\sim 11$), while the peak of the SNR distribution for the Phase 2 detections is $\sim 6$ (with median $\sim 7$). This significant improvement in detecting low SNR sources in Phase 2 from the source finding runs can mainly be attributed to the following reasons -- a) the overall data quality of the Phase 2 observations has improved significantly compared to Phase 1 data mainly because the fields targeted in Phase 2 were chosen specifically to avoid continuum sources brighter than 2 Jy. This leads to better data quality with fewer continuum-related artefacts, leading to the source finding runs being more complete out to low SNR; b) the on-dish calibrators were switched off for Phase 2, as they had caused a lot of RFI in the Phase 1 data, particularly in the corner beams; c) the SoFiA settings were fine-tuned based on the experience participating in the SKA Science Data Challenge 2 \citep{Hartley2023}, which has also contributed to a higher completeness of the catalogue in Phase 2.

Panel \textbf{c)} in Figure~\ref{fig:PDR2_properties} shows the distribution of the local rms noise in the image cubes for both Phase 1 and 2 sources. The median rms in the image cubes for Phase 2 is $\sim 1.7$ mJy per 30\arcsec\ beam and 18.5 kHz ($\sim 4 $ \kms) channel width, which is close to the expected theoretical rms noise in the image cube for WALLABY \citep{Koribalski2020}. This translates to a 5$\sigma$ \h1 column density (N$_{\textrm{\h1}}$) sensitivity of $\sim 9.1\times10^{19}(1 + z)^4$ cm$^{-2}$ per 30\arcsec\ beam and $\sim 20$ \kms channel, and a 5$\sigma$ \h1 mass sensitivity of $\sim 5.5\times10^8 (D/100$ Mpc)$^{2}$ M$_{\odot}$ for point sources, where $D$ is the Hubble distance to the source.

In terms of the line width of the detections in Phase 2, we show the distribution of the $w_{20}$ line-widths (defined as the spectral width corresponding to 20\% of the peak flux in the integrated spectrum) for both the Phase 1 and Phase 2 samples in panel \textbf{d)} in Figure~\ref{fig:PDR2_properties}. The median $w_{20}$ value for both the samples is $\sim 170$ \kms. 

Panel \textbf{e)} in Figure~\ref{fig:PDR2_properties} shows the distribution of the major axis size of the ellipse fit to the moment 0 map of the detections. It can be seen that the median size of sources detected in Phase 2 is $\sim 1.3$ beams, at the nominal 30\arcsec\ resolution, compared to a median value of $\sim 1.6$ for the Phase 1 detections. This means that WALLABY has managed to detect a larger number of marginally resolved galaxies in Phase 2, primarily because the median redshift of Phase 2 detections is a factor of two higher than the median value for Phase 1 observations. 

Panel \textbf{f)} in Figure~\ref{fig:PDR2_properties} shows the distribution of the \h1 mass for all pilot Phase 1 and 2 detections. The \h1 mass is computed using equation 7 in the PDR1 paper \citep{westmeier2022}. We observe that the Phase 2 detections have a median \h1 mass of $\log_{10}(M_{\textrm{\h1}}$/M$_{\odot}) \sim 9.6$ which is consistent with the median \h1 mass value of $\log_{10}(M_{\textrm{\h1}}$/M$_{\odot}) \sim 9.5$ for the pilot Phase 1 detections. The phase 2 median \h1 mass is slightly higher than the Phase 1 median \h1 mass, which is expected from the higher median redshift of the Phase 2 sample. We note that we make use of the Hubble distance,  $D = v H_{\textrm{0}}$, of the sources to estimate their \h1 mass. Where $v$ is the measured barycentric velocity and $H_{\textrm{0}} = 70$~\kms~Mpc$^{-1}$ is the Hubble constant. We caution that this distance is only an approximation and will be prone to large errors of up to $\sim 20$\% due to effects of peculiar velocities in the local Universe, as well as systematic errors from using barycentric redshifts (\citealt{Strauss1995};~\citealt{Willick1997}). We have used the Hubble distances for this release to remain consistent with the distance estimates used in Phase 1. However, going forward, for the full survey, the WALLABY team plans to apply more sophisticated flow models and correct the redshifts appropriately before measuring derived quantities such as distances and \h1 masses.

\begin{figure}
    \centering
    \includegraphics[width=1\columnwidth]{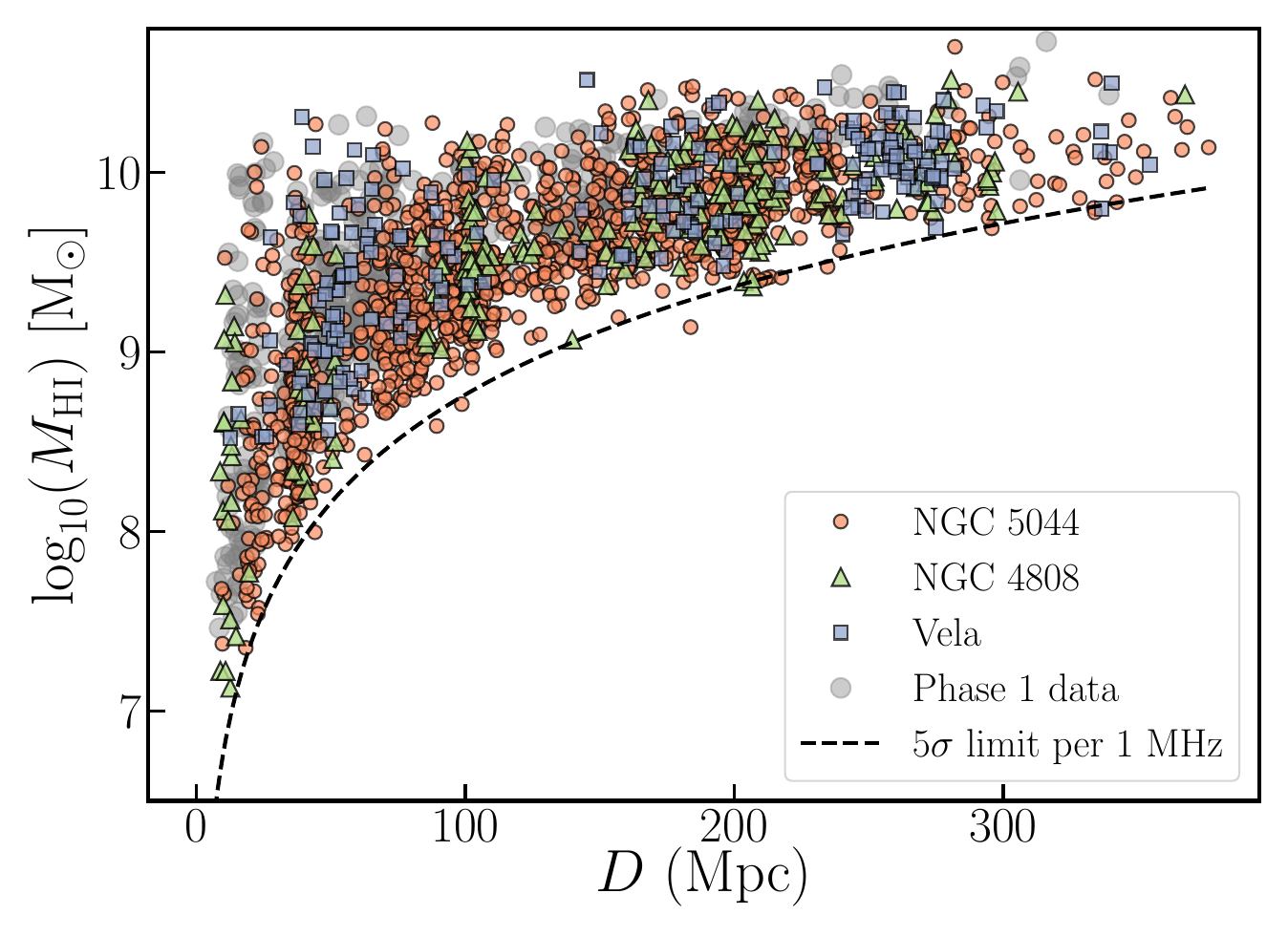}     
    \caption{The \h1 mass plotted against the estimated Hubble distance for the combined Pilot Phase 2 sample. The orange circles represent the NGC 5044 field, green triangles the NGC 4808 field and the purple squares the Vela field. The grey circles in the background represent the Phase 1 detections. The dashed black line represents the 5$\sigma$ \h1 mass threshold as a function of distance, assuming a 1 MHz frequency band width.}
    \label{fig:MHI-D_all}
\end{figure}

Figure~\ref{fig:MHI-D_all} shows the distribution of the \h1 mass of the detections from both pilot Phase 1 (grey points) and Phase 2 (color-coded by the different fields) as a function of their measured Hubble distance ($D = v H_{\textrm{0}}$). Also plotted is the 5$\sigma$ \h1 mass detection threshold (dashed black line) measured across a 1 MHz frequency bandwidth and assuming the median local RMS noise level of $\sim 1.71$ mJy in the image cubes derived from the SoFiA runs. The 5$\sigma$ \h1 mass detection threshold is computed as follows:
\begin{equation}
   \frac{M_{\textrm{\h1}} (5\sigma)}{M_{\odot}} = \frac{5 \times 49.7 \times \left(\frac{\sigma} {\textrm{\footnotesize Jy Hz}}\right)}{\sqrt{\Delta \nu/d\nu}} \left( \frac{D}{\textrm{Mpc}} \right)^2 
\end{equation}

Where, $\sigma = 1.71 \times 10^3$ Jy Hz and $\Delta \nu = 1000$ kHz is the 1 MHz channel width and $d\nu = 18.5$ kHz is the default spectral resolution. We see that our completeness at 5$\sigma$ is close to zero in accordance with Figure~\ref{fig:completeness} in Section~\ref{subsec:completeness}. As with the Phase 1 detections, we find large-scale clustering at various distances corresponding to the different groups detected in the Phase 2 fields. For example, for the NGC 4808 field, we find galaxies clustered at $\sim 30$ Mpc, $\sim 100$ Mpc and another over-density close to $\sim 200$ Mpc. Similarly, for the Vela field, we find an over-density of galaxies corresponding to a distance of $\sim 50$ Mpc, at $\sim 180$ Mpc and another at $\sim 260$ Mpc. The over-density at $\sim 260$ Mpc in Vela field is particularly interesting as it lies in the Zone of Avoidance (ZOA) and as such there are limited redshifts. However, a few previous optical studies (e.g.,~\citealt{Hudson2004};~\citealt{Hoffman2015}) have hinted at the existence of a large over-density corresponding to a systemic velocity of $\sim 18000$ \kms (roughly a distance of 260 Mpc). This was later confirmed by \cite{Kraan-Korteweg17}, who measured the spectra from $\sim 4500$ galaxies to map the composition and structure of the over-density. Studying and understanding this large-scale structure will add immensely to our knowledge of modelling bulk flows in the local Universe, as well as mapping the large-scale structures in the ZOA.

\subsection{Completeness}
\label{subsec:completeness}

\begin{figure}
    \centering
    \includegraphics[width=\columnwidth]{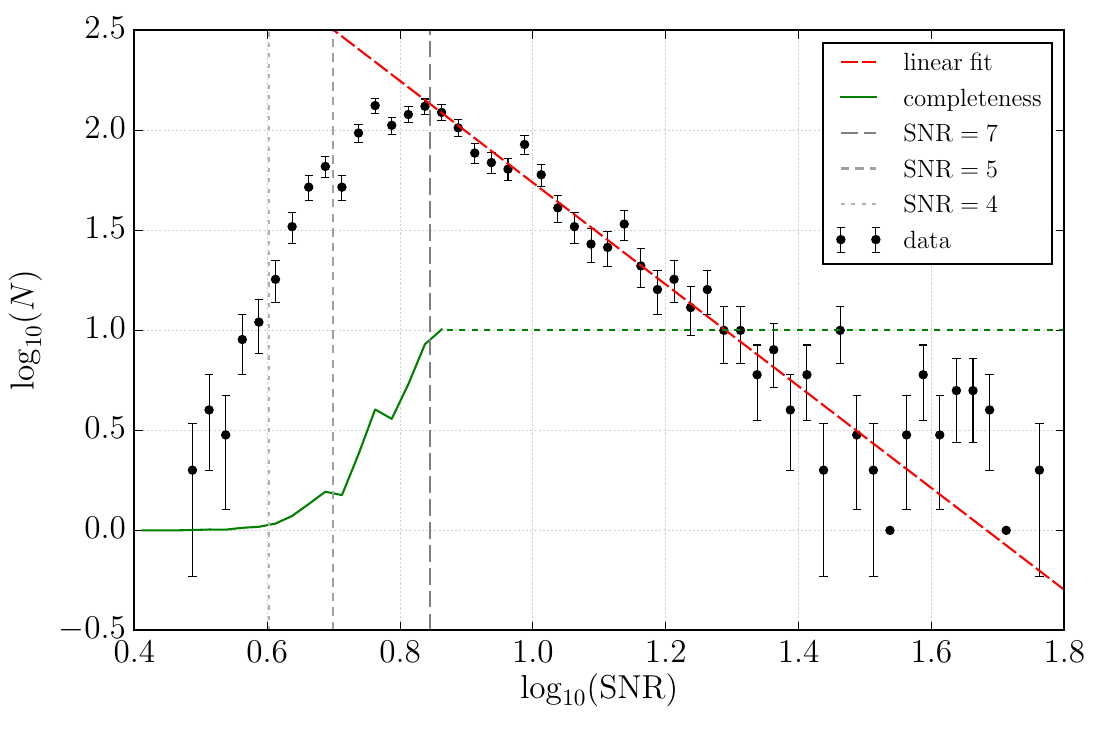}
    \caption{Histogram of the number of detected sources, $N$, as a function of integrated signal-to-noise ratio, SNR, in double-logarithmic space in bins of $\Delta \log_{10}(\mathrm{SNR}) = 0.025$ (black data points). The error bars correspond to $\sqrt{N}$. The red, dashed line shows the result of a linear fit in the range of $0.9 < \log_{10}(\mathrm{SNR}) < 1.4$. The resulting completeness, defined as the observed source count divided by the fit, is shown as the green, solid curve at $\mathrm{SNR} \lesssim 7$ where incompleteness effects are evident.}
    \label{fig:completeness}
\end{figure}

In order to estimate the completeness of the source catalogue, we plot in Figure~\ref{fig:completeness} the number of sources, $N$, as a function of integrated signal-to-noise ratio (SNR) in double-logarithmic space. As before the SNR is defined here as the ratio of the integrated flux and the statistical uncertainty of the integrated flux measurement within the source mask produced by SoFiA. As expected from an untargeted survey, the source count follows an almost perfect power-law with a turnover at $\mathrm{SNR} \lesssim 7$. Under the assumption that the intrinsic population continues to follow a power law at low SNR and that the turnover therefore is entirely caused by incompleteness, we can estimate completeness as a function of SNR. We do this by fitting a straight line to the data points in the range of $0.9 < \log_{10}(\mathrm{SNR}) < 1.4$ (red, dashed line) which yields a slope, and hence power-law exponent, of $-2.54$. The completeness of our source catalogue as a function of SNR can then be estimated by dividing the number of detected sources in each bin by the expected number of sources predicted by the power-law fit.

The resulting completeness curve is shown as the green, solid line in Figure~\ref{fig:completeness}. We reach 100\% completeness at $\mathrm{SNR} \approx 7$ beyond which we do not plot the actual completeness curve any more, as it would eventually show a large scatter around $1$ due to stochastic errors as a result of low source counts at high SNR. 50\% completeness is reached at $\mathrm{SNR} \approx 5.5$ below which our completeness rapidly declines to near zero at $\mathrm{SNR} \approx 4$.

%%%%%%%%%%%%%%%%%%% Section: High-resolution 12 arcsec "cut-outs"     %%%%%%%%%%%%%%%%%%%%%%%%%%%%%%%%%%%%%%%%%%%%%
\section{High-resolution 12\arcsec ~cut-outs}
\label{sec:postage_stamps}

One of the objectives of the WALLABY survey is to generate high-resolution (12\arcsec) cut-outs for a sub-sample of galaxies. We use the calibrated visibility data derived from the default ASKAP spectral-line processing pipeline (\citealt{Guzman2019};~\citealt{Whiting20}) to image a sub-sample of galaxies at high angular resolution. As mentioned earlier the default spatial resolution of the WALLABY survey is 30\arcsec, which was determined to be the optimal resolution that gives a good compromise between resolution, sensitivity and computational resources required to process large volumes of data. In contrast, the computational resources required to image the data in the full 12\arcsec\ resolution will be significantly higher due to the additional baselines and increasing image sizes. However, it is still possible to image a sub-sample of the WALLABY detections in high-resolution by limiting the bandwidth to be imaged to a few hundred channels and only encompassing the velocity range of the target galaxies. This way, we drastically reduce the computing and storage requirements to process the data. We tested this functionality in preparation for the full WALLABY survey in Phase 2.

For Phase 2, we selected all HIPASS sources from the three fields. We targeted HIPASS sources, as these are likely to be detected in the WALLABY data and also as they are well resolved (tens of 12\arcsec\ beams across the major axis). We note that for the full WALLABY survey, apart from the HIPASS targets, some optically-selected target galaxies are also expected to be included. We emphasise here that since the target galaxies for the high-resolution cut-outs are HIPASS galaxies and therefore \h1-selected, this will naturally introduce biases in the sample, which the users need to consider and account for while using the data for their analysis. 

To perform the high-resolution imaging making use of the full visibility including the longest baselines, we split out individual ASKAP primary beams containing (and surrounding) our target sources. We split out 250 channels ($\sim$4.6 MHz) encompassing the velocity range of the source. For the WALLABY channel width of $\sim 4$ \kms, this translates to a total velocity range of $\sim 1000$ \kms, which is sufficient to contain the emission from even the most massive and rotationally-dominated galaxies. We split out only 250 channels mainly to bring down the storage and processing costs required for each source. We split out up to 3 PAF beams from each footprint for each source, i.e., up to a total of 6 beams for a single source from both footprints. Each calibrated visibility data set of 250 channels for each beam is $\sim 15$GB in size, therefore the total storage cost for each source for 6 beams is $\sim 90$GB. The split-out visibilities are then uploaded on to CASDA. The splitting of the visibilities described above is performed automatically whenever a new field has been observed and processed. 

The relevant visibilities for each source are then downloaded from CASDA and used to make the high-resolution image cubes using the ``high-resolution'' imaging pipeline (hereafter high-res pipeline). All data have been reduced on Pawsey Supercomputing Facility's dedicated High Performance Computing clusters. We make use of ASKAPSoft to process and image the cut-outs. We now describe the various stages of the high-res pipeline. The pipeline is a Python script that reads in a catalogue of sources that need to be imaged, and a user-defined configuration file containing essential information such as the location of the split-out calibrated visibility, holography and footprint data. The main Python pipeline job then creates all the necessary bash scripts such as the parsets and the corresponding slurm job submission scripts for each task (e.g., imager, imcontsub etc). These jobs for the various tasks are then submitted as dependencies for each beam for each individual source in a parallel framework. 

The imaging is first carried out beam-by-beam and then all beams are mosaicked to produce the final image cube for the individual sources. The first step is to image the visibilities for each beam using the \texttt{cimager} task in ASKAPSoft. We make an image of size 384 $\times$ 384 pixels, with a pixel size of 2\arcsec. We use a Wiener filter with a robust parameter value set to 0.5 and apply a Gaussian taper of 12\arcsec\ to achieve a synthesized beam close to 12\arcsec. The spectral resolution is kept at 18.5 kHz ($\sim 4$~\kms). In addition, the deconvolution process is also performed within the task \texttt{cimager}. For more details on the ASKAPSoft parameters used for the imaging, refer to Table~\ref{tab:imaging_params} in \ref{appendix:imaging_params}. The imaging step is then followed by the image-based continuum subtraction using the task \texttt{imcontsub}. The pipeline then performs the primary beam correction using the holography model with the task \texttt{linmos}. These steps are performed for each of the 6 beams that encompass the target HIPASS source. As the final step, all 6 beams are mosaicked to form the final `mosaicked' cube for the source. This is again performed using the mosaicking task \texttt{linmos}. The above workflow is adopted for all sources and a number of jobs are submitted on the cluster to simultaneously image the data for multiple sources at any given time. We now present an overview of the cut-outs sample, and give details of the quality of the data, including the typical SNR of the detections, size distribution and their \h1 mass range. In addition, we also compare the properties of the 12\arcsec\ detections with their corresponding 30\arcsec\ counterparts.

\subsection{12\arcsec\ imaging results}
\label{subsec:12arc_results}

\begin{figure*}
    %\centering
    \vspace*{-0.75cm}
    \includegraphics[width=1\textwidth]{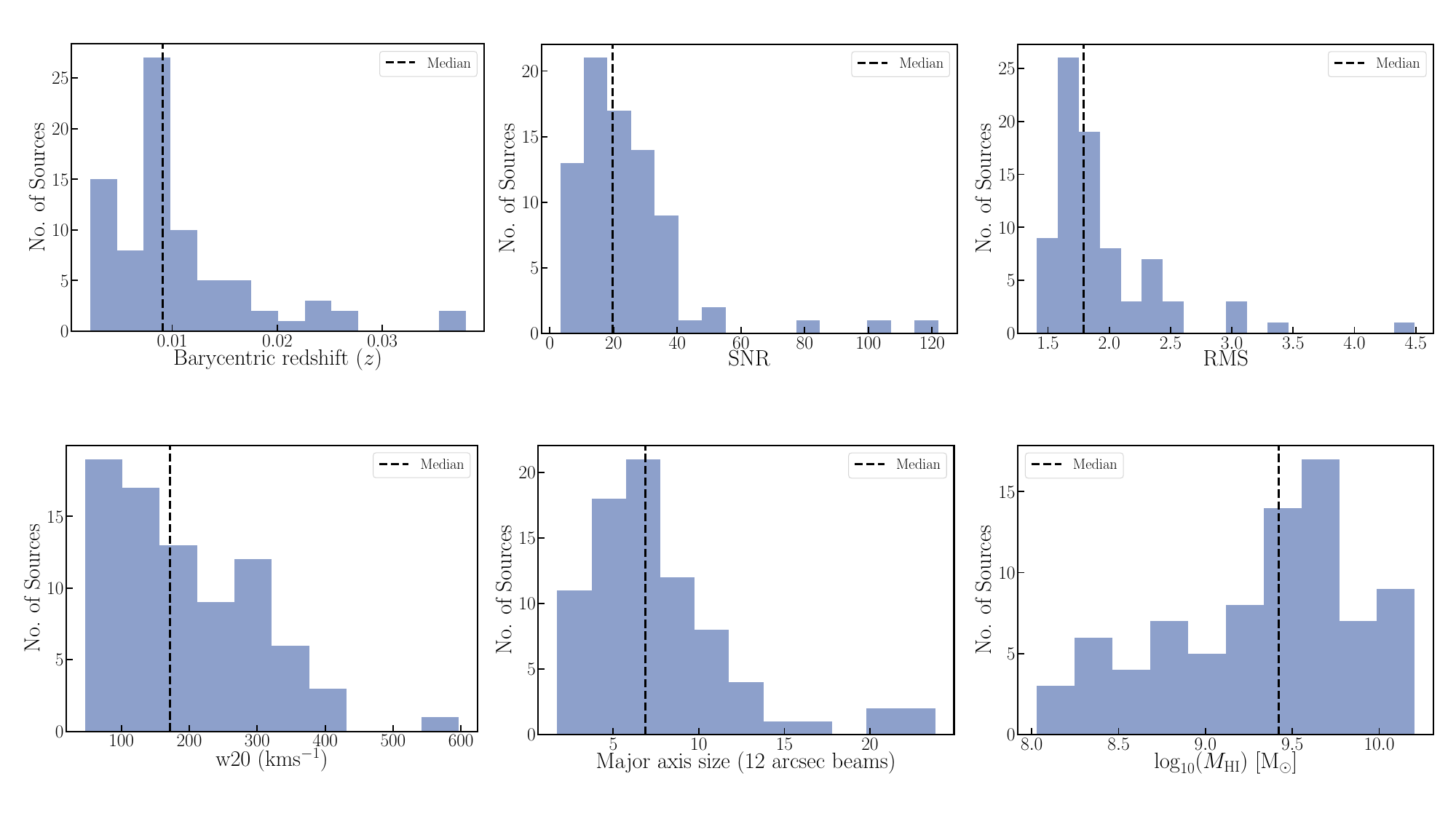}     
    \caption{Plots show the source properties of the 12\arcsec\ detections in the Phase 2 sample. \textit{Top left:} Distribution of the barycentric redshifts of the 12\arcsec\ detections. \textit{Top centre:} Histogram of the Signal-to-noise (SNR) of the 12\arcsec\ detections. \textit{Top right:} Local rms noise distribution in the images cubes. \textit{Bottom left:} Distribution of the $w_{20}$ \h1 line-width distribution. \textit{Bottom centre:} Histogram of the major axis size (in units of 12\arcsec\ beams). \textit{Bottom right:} The \h1 mass distribution. In all plots, the dashed black line represents the median value of the distribution. }
    \label{fig:12arc_properties}
\end{figure*}

A total of 73 HIPASS target galaxies were imaged in high-resolution as part of the Pilot Survey Phase 2. We note that in the majority of cases the target HIPASS galaxy is the only genuine detection in the image cube. However, in a few cases, source finding on some target HIPASS image cubes resulted in the detection of genuine smaller sources surrounding the target HIPASS galaxy. Once the source finding is complete, each tentative detection is visually examined to verify if it is a genuine source and is then added to the final source catalogue. A total of 80 sources are detected from the source finding runs from all three Phase 2 fields combined. Most detections in the high-resolution image cubes are also detected in the default 30\arcsec\ data cubes, however, in some cases it is observed that a 30\arcsec\ source in the default WALLABY catalogue is split-up into multiple components, with each component being a genuine nearby galaxy in the vicinity of a large galaxy. In such cases, each 12\arcsec\ component is assigned a unique WALLABY name.

In Figure~\ref{fig:12arc_properties}, we plot some of the source characteristics of the 12\arcsec\ sample. Given that most targeted 12\arcsec\ sources are HIPASS detections, the redshift distribution of the sample ranges from $0.002 < z < 0.04$, with a median $z \sim 0.01$. We find that the median SNR of the 12\arcsec\ detections is $\sim 20$, while the rms in the local image cubes of the 12\arcsec\ detections is found have a median value of $\sim 1.8$ mJy, which is close to the expected theoretical rms of 1.75 mJy (using robust=0.5 and all baselines including 6 km). This translates to a 5$\sigma$ \h1 column density (N$_\textrm{\h1}$) sensitivity limit of $\sim  6\times10^{20}(1 + z)^4$ cm$^{-2}$ assuming a 12\arcsec\ beam and a 20 \kms channel width, which is a factor of 6.6 higher compared to the 30\arcsec\ data. This is a natural compromise between sensitivity and spatial resolution that is associated with higher-resolution observations and we advice the user to be cognisant of this compromise in sensitivity when dealing with the high-resolution data.

We also note that the distribution of the $w_{20}$ \h1 line-width for the 12\arcsec\ detections ranges from $46 < w_{20}\ (\textrm{\kms}) < 597$, with a median $w_{20} \sim 172$ \kms, which indicates that the majority of the high-resolution sources are likely to be rotationally-supported late-type galaxies. We examined the moment maps and the corresponding optical image for the obvious outlier (with $w_{20} \approx 597$ \kms) and find that the \h1 emission is much more extended compared to the optical disk, along with kinematic warps and other signatures indicating that this galaxy is likely undergoing an interaction and may have accreted \h1 gas from a gas-rich low-mass companion. As the SoFiA mask encompasses all the \h1 emission, it results in considerably broadening the velocity width of this detection. Most of the 12\arcsec\ detections are well resolved with their major axis size typically spanning $\sim 7$ (12\arcsec) beams across. Finally, we note that the \h1 mass distribution of the high-resolution sample is $8.0 \leq \log_{10}\left(\frac{M_{\textrm{\h1}}}{\textrm{M}_{\odot}} \right) \leq 10.2$, with a sample median of $\log_{10}\left( \frac{M_{\textrm{\h1}}}{\textrm{M}_{\odot}} \right) \sim 9.42$. 

\begin{figure*}
    %\centering
    \vspace*{-0.3cm}
    \includegraphics[width=1\textwidth]{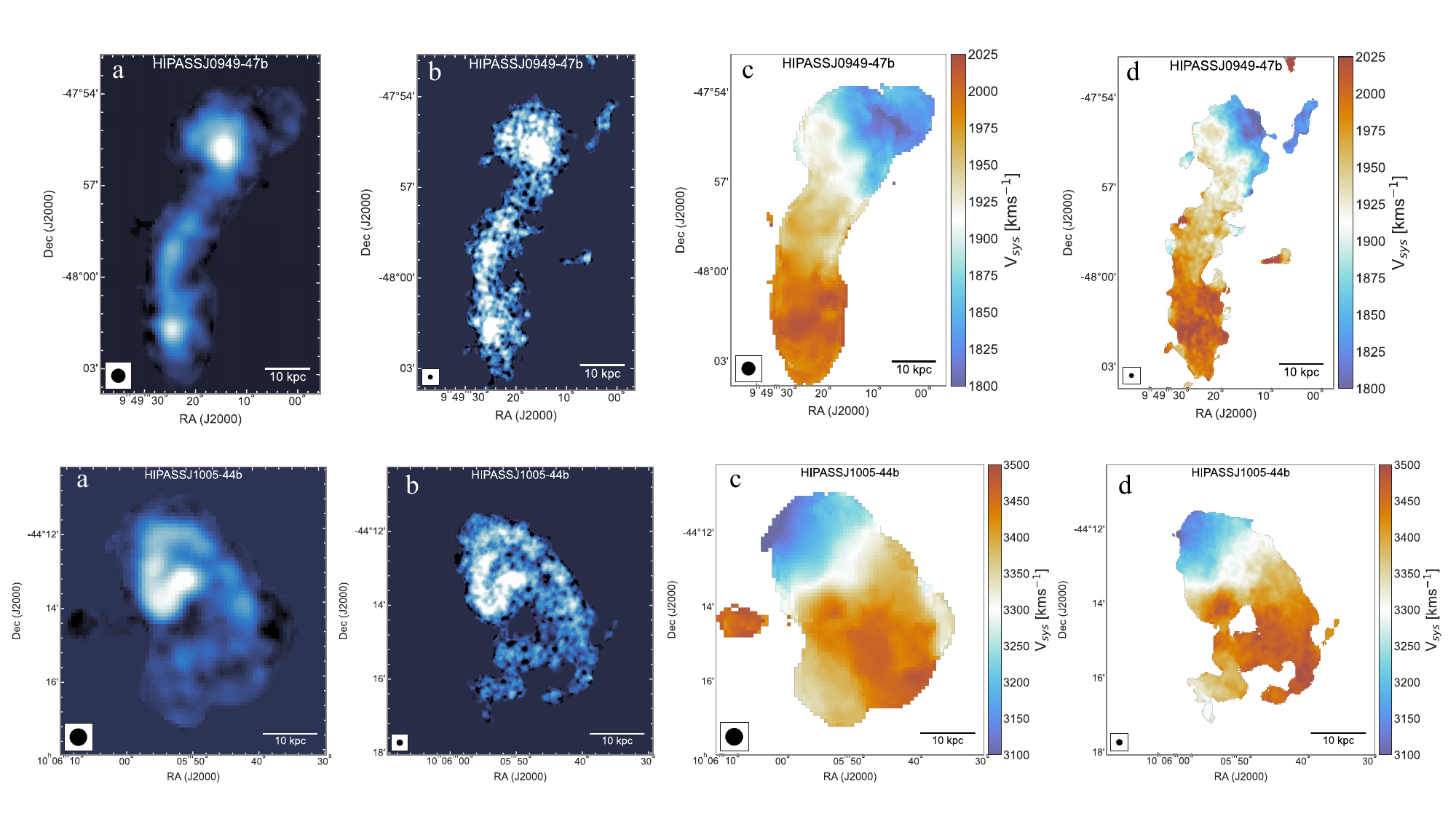}
    \caption{The comparison of moment 0 and moment 1 maps for two galaxies (top: HIPASS J0949-047b, bottom: HIPASS J1005-44b) with a resolution of 30\arcsec\ and 12\arcsec. In each row, panels (a) and (c) show the moment 0 and 1 maps with a resolution of 30\arcsec\, while panels (b) and (d) show the corresponding 12\arcsec\ maps. At the bottom of each figure, we show the respective beam size as black circles and a scale bar set to 10 kpc.}
    \label{fig:sample_mosaic1}
\end{figure*}

\begin{figure*}
    \centering
    \includegraphics[width=1\textwidth]{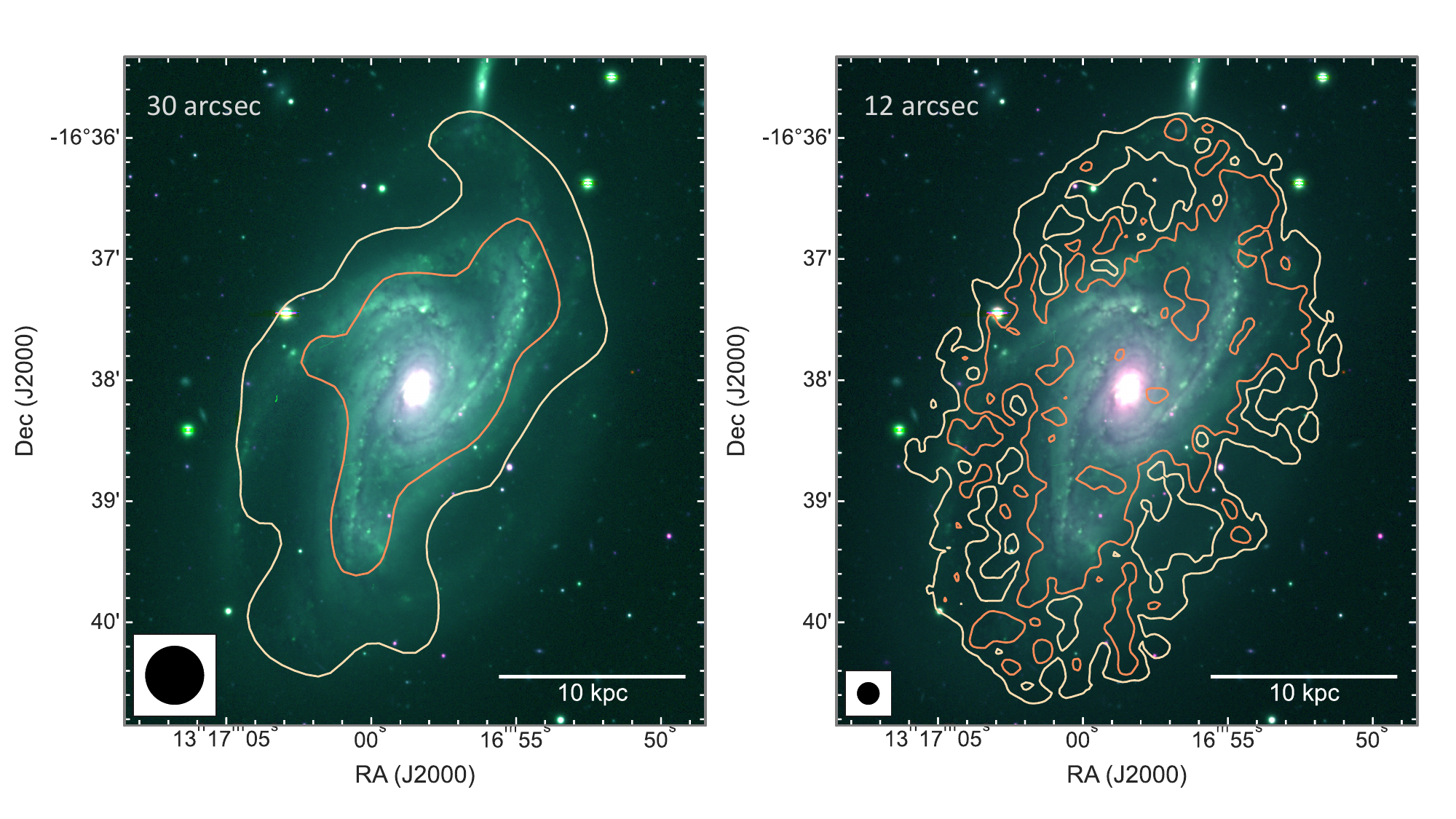}
    \caption{\textit{Left:} 30\arcsec\-resolution \h1 contours overlaid on top of a composite (g,z,i) DESI Legacy Survey image of the galaxy NGC 5054. \textit{Right:} Corresponding 12\arcsec\-resolution \h1 contours. In both cases the contours levels are set at column densities of 2.4$\times 10^{20}$ cm$^{-2}$ (light orange) and 7.2$\times 10^{20}$ cm$^{-2}$ (dark orange).}
    \label{fig:NGC5054_desi_HI_overlay}
\end{figure*}

Figure~\ref{fig:sample_mosaic1} shows moment 0 (intensity) and 1 (velocity) maps of two interacting system of galaxies in the default 30\arcsec\ and 12\arcsec-resolution. From the images it is very clear that finer details in the \h1 morphology begin to show-up in the high-resolution images. The high-resolution moment maps highlight the distribution of the high-column density \h1 gas in the galaxies, which are otherwise washed-out in the 30\arcsec\ images. In addition, in Figure~\ref{fig:NGC5054_desi_HI_overlay} we show the 30\arcsec\ and 12\arcsec\-resolution \h1 contours overlaid on top of a composite (g,z,i-band) DESI Legacy Survey image for the galaxy NGC 5054. The two contours show \h1 column densities of $2.4 \times 10^{20}$ cm$^{-2}$ (light orange) and $7.2 \times 10^{20}$ cm$^{-2}$ (dark orange), respectively. The contours correspond to a SNR of 4 and 10 in the 30\arcsec\ image respectively, while corresponding to a SNR of 2 and 6 in the 12\arcsec\ image. Compared to the 30\arcsec\ resolution \h1 contours, the 12\arcsec\ resolution contours clearly trace the high-column density \h1 gas along the spiral arms in NGC 5044, allowing us to study both the \h1 gas and star formation properties at a much higher resolution. A factor of $\sim 3$ improvement in resolution will significantly aid in studies directed towards understanding the distribution of the high-column density gas in galaxies and also enable us to more accurately probe the connection between \h1 gas, star formation and star formation laws. In addition, the higher resolution enables us to model the kinematics of the \h1 gas more accurately.

%%%%%%%%%%%%%%%%%%%%%%%%%%%%%%%%%  Subsection: Data quality and known issues with the high-resolution data  %%%%%%%%%%%%%%%%%%%%%%%%%%%%%%%%
\subsection{Data quality and known issues with the high-resolution data}

We do note that while the overall quality of the 12\arcsec\ data is good, there were some issues identified with the imaging pipeline as well as the data products. We list below some of the known issues with the cut-outs in this data release. \\

\noindent \textbf{Flux discrepancy:} We note that the 12\arcsec\ sources show a higher integrated flux compared to their 30\arcsec\ counterparts. The flux of the 12\arcsec\ sources is on average $\sim$15\% higher compared to their 30\arcsec\ counterparts. We present a more thorough discussion on this flux discrepancy in Section~\ref{sec:flux_discrepancy} and also highlight the likely origins of the discrepancy. \\
 
\noindent \textbf{Different synthesized beam size:} Some sources from the NGC 5044 field (tile4) have a different angular resolution. These data sets have a synthesized beam of $\sim$17\arcsec\ instead of 12\arcsec. There are 14 such sources. This is because a slightly different tapering was applied during the imaging stage. The visibilities for these sources were not stored as the observations for the NGC 5044 tile 4 were carried-out before the scheme of storing visibilities on to CASDA was introduced. As such, the visibilities for these sources were unfortunately unavailable to be re-imaged  to a 12\arcsec\ resolution. We have included a comment in the source catalogue for all relevant affected sources to highlight this. \\

\noindent \textbf{Unreliable spectra:} 7 sources in the 12\arcsec\ data show bad spectra. These are typically edge-on galaxies with large spectral widths. Given that only 250 channels are split-out for the high-resolution imaging, we suspect that there were not enough line-free channels for the image-based continuum subtraction routine in ASKAPSoft to properly perform the continuum subtraction, leading to over-subtraction. Sources affected by this issue have a \textbf{qflag = 128} in the source catalogue. \\

\noindent \textbf{No default 30\arcsec\ WALLABY cross-match:} We note that 6 sources in the cut-outs source catalogue do not have a corresponding default 30\arcsec\ WALLABY detection. Upon further examination, it was found that the missing sources in the 30\arcsec\ WALLABY catalogue are due to one of the following reasons.
\begin{itemize}
    \item Source lies in the Galactic velocity range. The default 30\arcsec\ source finding runs are only performed on the extra-galactic velocity range ($cz \sim 500 - 26500$ \kms) and as a consequence all sources below a velocity of $cz < 500$~\kms\ are excluded from the source finding runs. Two sources are missing due to this limitation.
    
    \item Source is in the corner of a footprint. The SoFiA source finding runs are only performed on the inner $4^{\circ} \times 4^{\circ}$ area of the mosaicked footprint as the outer edges of the footprint suffer from lower SNR and sensitivity as the noise increases by a factor of two. For this reason, some sources in the outer parts of the specific footprint may have been omitted in the current default 30\arcsec\ source finding run. These sources will however be added to the catalogue whenever overlapping footprints are subsequently processed and available for source finding. Three sources are missed due to this.
    
    \item Very faint sources near the detection threshold may be missed in the global 30\arcsec\ source finding, as the completeness curve is known to gradually decrease below an SNR of $\sim 7 - 8$ (see Figure~\ref{fig:completeness}). Since the high-resolution source finding involves checking and verifying each individual detection, in some cases it is possible to detect sources close to the detection threshold of the source finding runs. One source is missed due to this issue.
\end{itemize}

%%%%%%%%%%%%%%%%%%%  Section: Flux Accuracy  %%%%%%%%%%%%%%%%%%%%%%%%%%%%%%%%%%%%%%%%%%%%%%%%%%%%%%%%%%%%%%%%%%
\section{Flux discrepancy}
\label{sec:flux_discrepancy}

\begin{figure}
    %\centering
    \vspace*{-3.2cm}
    \hspace*{-0.7cm}
    \includegraphics[width=1.25\textwidth]{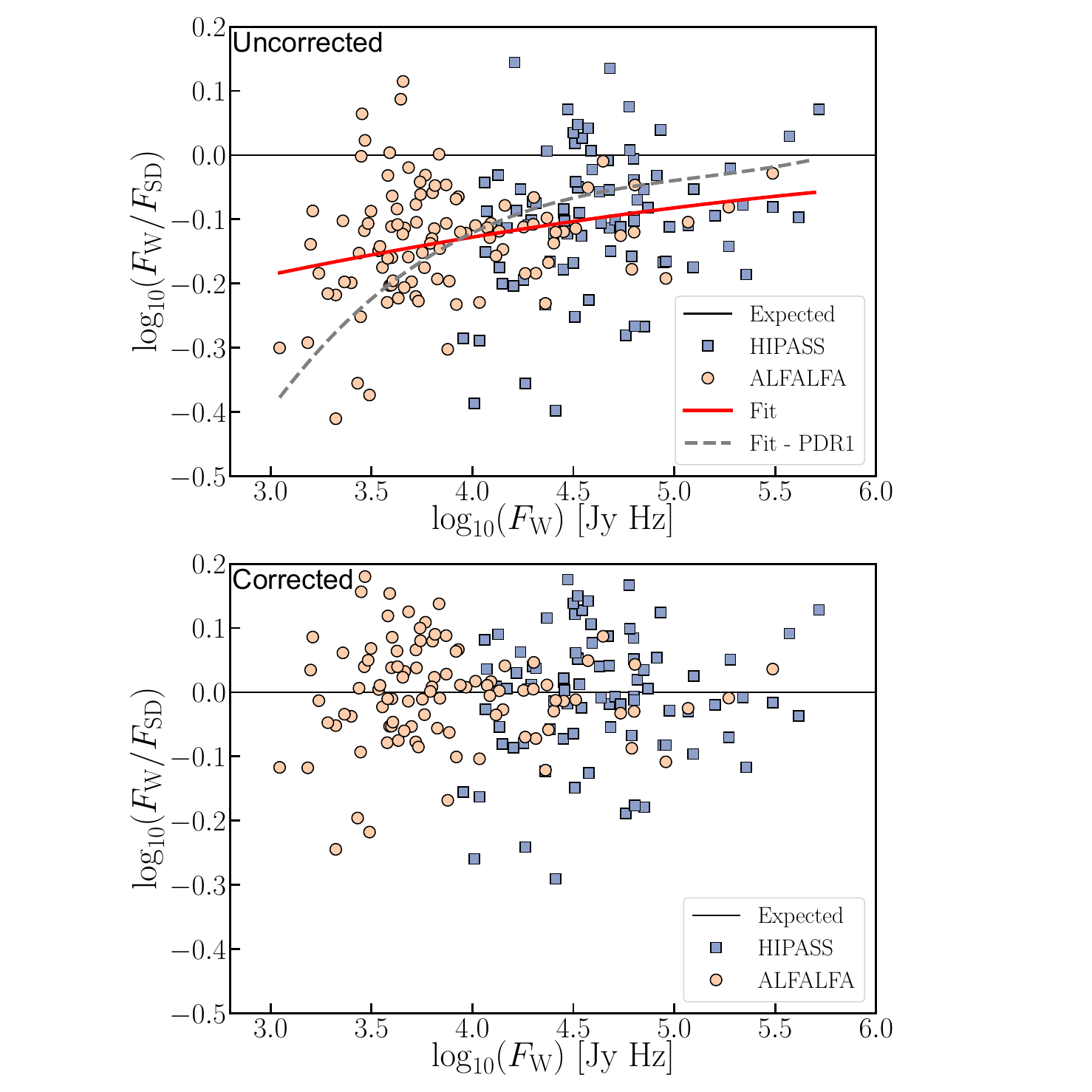}     
    \caption{\textit{Top:} The ratio of the WALLABY 30\arcsec\ integrated flux to the single-dish integrated flux plotted against the WALLABY integrated flux for those galaxies which have a corresponding single-dish cross-match, either in ALFALA and/or HIPASS. For the NGC 5044 and Vela fields, we use the HIPASS data and for the NGC 4808 field, we use the ALFALFA data for the flux comparison. \textit{Bottom:} Similar plot as above, but now the WALLABY fluxes have been corrected using a polynomial fit to the data. The horizontal black line represents a flux ratio of one in both cases.}
    \label{fig:flux_compare_30arcsec}
\end{figure}

\begin{figure}
    %\centering
    \vspace*{-2cm}
    \hspace*{-0.72cm}
    \includegraphics[width=1.3\textwidth]{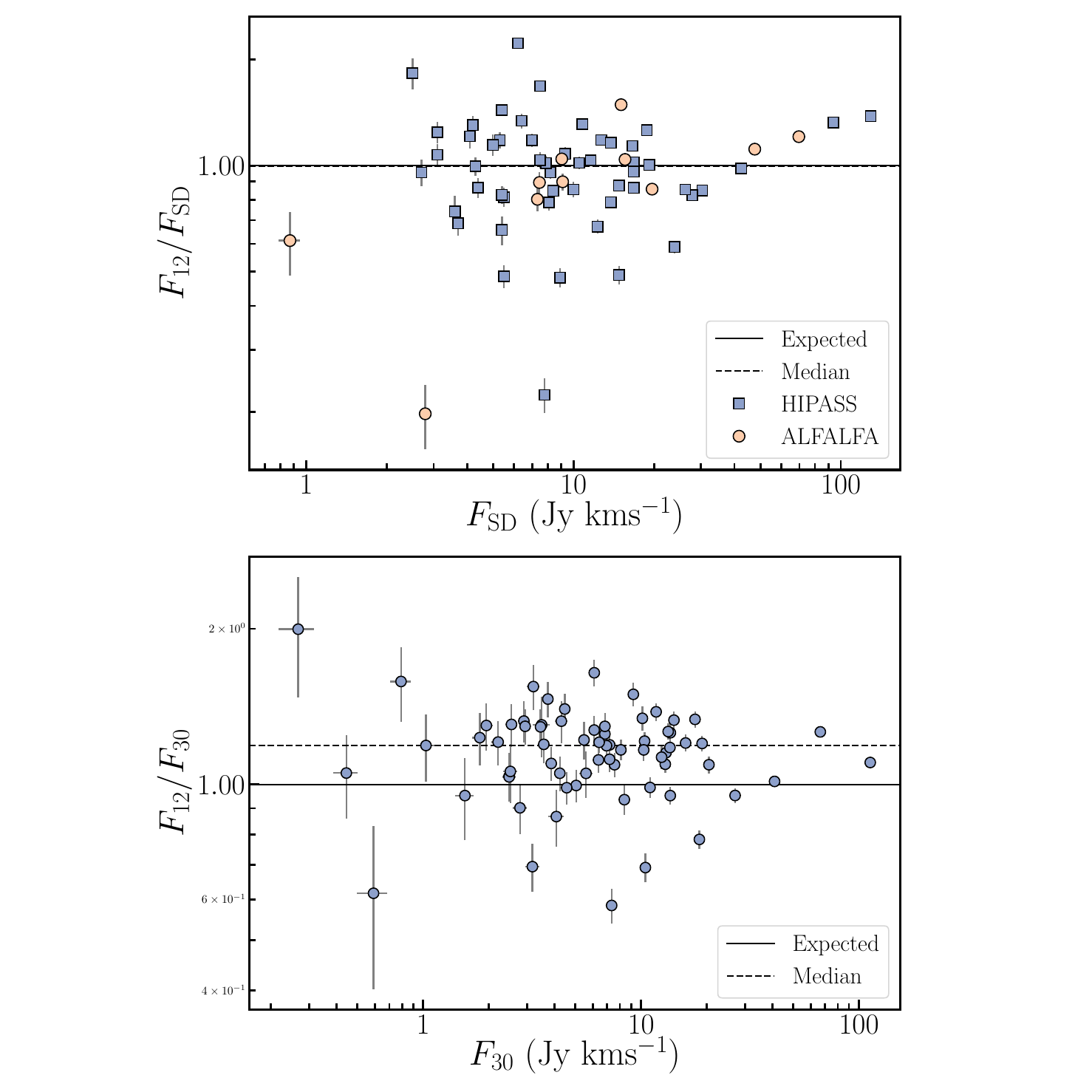}
    \caption{Plot shows the ratio of the integrated flux of the 12\arcsec\ ($F_{12}$) to the 30\arcsec\ flux ($F_{30}$) for the overlapping sample. The black solid line represents the expected one-to-one line, and the dashed black line represents the median value of the $\frac{F_{12}}{F_{30}}$ ratio.}
    \label{fig:flux_compare_12Vs30}
\end{figure}

As mentioned in the previous sections, the 30\arcsec\ detections are observed to have a flux deficiency of $\sim 15$\% compared to single-dish observations (see upper panel in Figure~\ref{fig:flux_compare_30arcsec}). The flux discrepancy between the 30\arcsec\ detections and their single-dish observations was also observed in the Phase 1 release. A number of independent effects may be contributing to this issue including flux that is genuinely missed in the 30\arcsec\ data from extended diffuse emission; inadequate deconvolution threshold, which leads to contributions from uncleaned flux and negative side lobes associated with the 30\arcsec\ dirty beam (see \citealt{westmeier2022}); as well as systematic flux offsets due to the different procedures implemented for the primary beam correction between Phase 1 and Phase 2. While it is difficult to estimate the typical fraction of flux that is missed from diffuse emission, it is easier to resort to simulations of ASKAP observations in order to specifically understand the impact of inadequate deconvolution thresholds on the measured flux. We discuss this in more detail in Section~\ref{subsec:Simulations}.

In addition, we observe a flux discrepancy between the 12\arcsec\ and the 30\arcsec\ sources, wherein the 12\arcsec\ detections typically show $\sim 15$\% higher flux compared to their corresponding 30\arcsec\ detections (see lower panel in Figure~\ref{fig:flux_compare_12Vs30}), while being consistent (albeit coincidentally) with their single-dish detections (see upper panel in Figure~\ref{fig:flux_compare_12Vs30}). We examined the ASKAP dirty beams for the 12\arcsec\ and 30\arcsec\ observations and find that there are significant positive sidelobes associated with the 12\arcsec\ beam, whereas there are significant negative sidelobes observed in the 30\arcsec\ dirty beams (see Figs. \ref{fig:12arc_PSF_sims} and \ref{fig:30arc_PSF_sims} in \ref{appendix2}). This observation is true for all declination ranges (-47\deg\ $ < \delta < $ +8\deg) covered by Phase 2 and could potentially lead to flux offsets, as any uncleaned flux in the residual maps impacted by the sidelobes of the dirty beams will contribute to the final image cube. In order to study the impact of the side lobes of the dirty beams for the two resolutions we resorted to simulations of ASKAP observations. In the subsection below we describe the details of the simulations and their main outcomes.

\subsection{Simulating ASKAP observations}
\label{subsec:Simulations}

In order to generate mock ASKAP observations, we make use of the MIRIAD software package \citep{Sault1995}. We use the MIRIAD task \texttt{UVGEN} to generate the mock visibilities. \texttt{UVGEN} takes in details of the mock observations, such as the positions of the ASKAP antennas, the correlator setup, the frequency of the observations, the RA and Dec of the pointing, hours of observation, integration time, the latitude of the observatory and the system temperature of ASKAP. This then generates the expected visibilities. We note that we generated mock visibilities at six different declinations (+8, +2, -11, -19, -24 and -45$^{\circ}$) to represent the declination range observed in the Phase 2 fields. We generated 4$\times$8 h mock ASKAP observations to approximately emulate the ASKAP beams which are processed and imaged independently before mosaicking them to form the final image cube with an rms noise close to $\sim 1.6$ mJy per beam per channel, representative of the noise in typical WALLABY image cubes. 

We Fourier-transform the mock visibilities to generate the dirty beam and dirty images at the 12\arcsec\ and 30\arcsec\ resolutions, using the MIRIAD task \texttt{INVERT}. We note that we were unable to yield a synthesised beam $\sim 12$\arcsec\ using a robust value of 0.5 and thus set the robust weighting to 0 and applied appropriate Gaussian tapering to the mock visibility data in order to achieve nearly 12\arcsec\ and 30\arcsec\ dirty beams. Next, we generate $\sim60$ mock galaxies per declination range, of varying size, surface densities, integrated flux, position angle and inclination angles using \texttt{BBarolo}'s \citep{diTeodoro15} \texttt{GALMOD} task. The size and the flux range of the model galaxies is set appropriately to reflect the respective range observed in the real 30\arcsec\ and 12\arcsec\ WALLABY detections. The model galaxy is then convolved with the 12\arcsec\ and 30\arcsec\ dirty beams, respectively, using the MIRIAD task \texttt{CONVOL}. The convolved galaxy models are then injected into the ASKAP observations dirty image cube (noise cube) using the task \texttt{MATHS}. This is then followed by the deconvolution step using MIRIAD's \texttt{CLEAN} task. We set the clean thresholds to match the settings used in the default ASKAPSoft imaging pipeline. This corresponds to a clean threshold in the minor cycle of 3.5 mJy ($\sim2\sigma$) and an additional second deep clean threshold of 0.5 mJy ($\sim0.3\sigma$). We set the maximum number of iterations (\texttt{niters}) for the minor clean cycles to be 800, again replicating the ASKAPSoft pipeline settings.

After the cleaning stage, we restore the images using the task \texttt{RESTOR}. This takes in the dirty beam, the dirty image (with the injected model galaxy) and the clean components from the deconvolution step and generates the residual map as well as the final restored image cube by convolving the clean components with 12\arcsec\ and 30\arcsec\ Gaussian beams, respectively, and adding them to the residual map. We then mosaic all four simulated image cubes to generate the final image cube on which source finding is performed using SoFiA.

In addition, we also convolved the model galaxies with a Gaussian of full width at half maximum (fwhm) of 12\arcsec\ and 30\arcsec\ which will represent perfectly ``cleaned'' data. The fluxes derived from these Gaussian beam convolved data sets will not have the influence of the sidelobes that is typically observed in interferometric data sets where some uncleaned flux may remain, impacting the final measured flux as well as the quality of the final image cubes. 

We run SoFiA on the simulated 30\arcsec\ image cubes by using the default parameter settings currently used for the WALLABY source finding. We appropriately change some SoFiA parameters for the high-resolution data sets as the default 30\arcsec\ parameters are not optimal for source finding in the 12\arcsec\ image cubes. For a list of important SoFiA parameters used for source finding on the 30\arcsec\ and 12\arcsec\ data-sets we refer the reader to Table~\ref{tab:SoFiA_params_30and12} in \ref{appendix:SoFiA_params}.

\subsection{Origin of the flux discrepancy}
\label{subsec:flux_issue_origins}

We now report the main observations from our simulation experiment and delve into the details of the origin of the flux discrepancy that is observed between the single-dish, 30\arcsec\ and 12\arcsec\ detections. 

\begin{figure*}
    \hspace*{-0.3cm}
    \includegraphics[width=1.05\columnwidth]
    {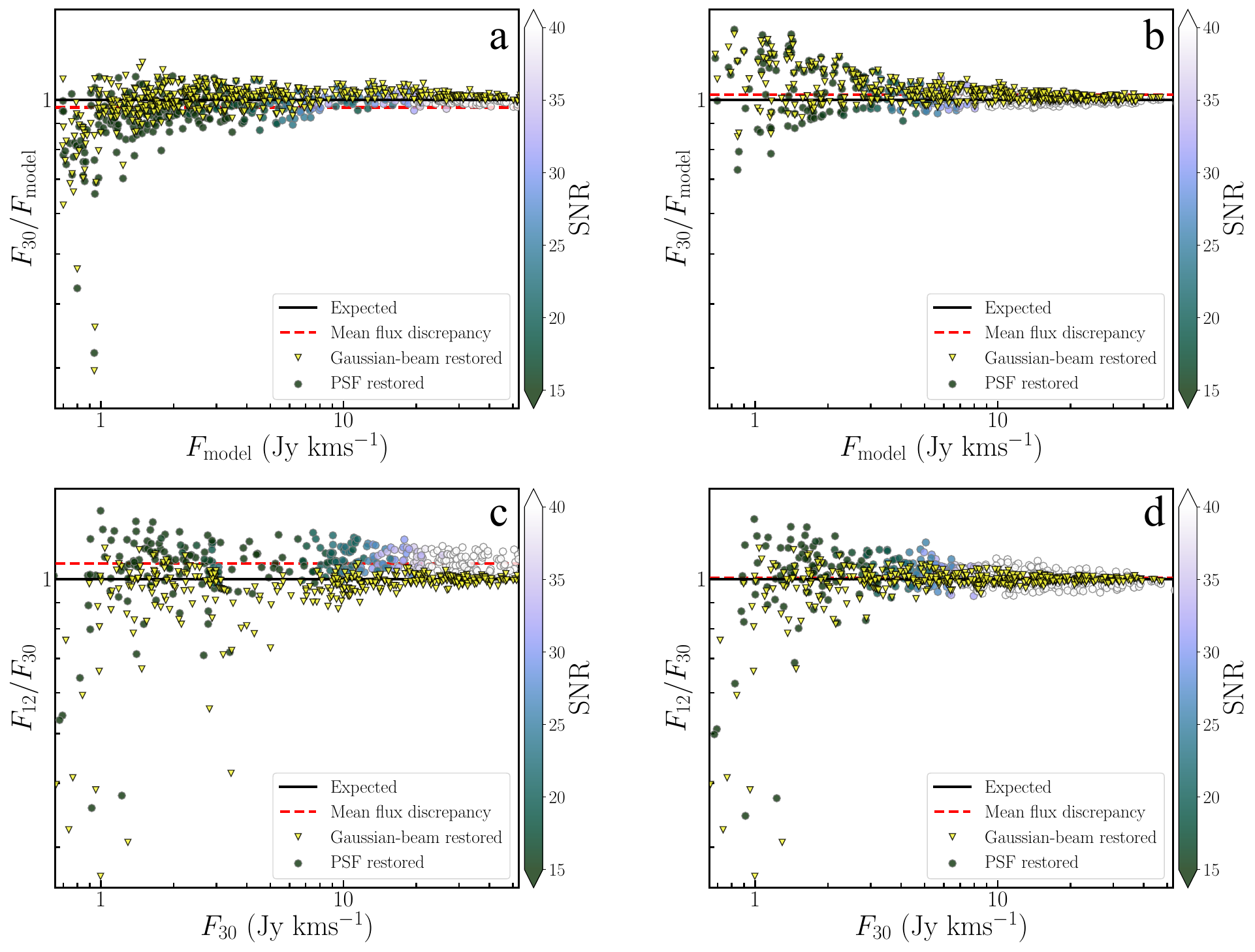}  
    \caption{\textbf{a):} Circles show the ratio of the integrated flux of the injected model source convolved with the 30\arcsec\ PSF ($F_{\mathrm{\small 30}}$) to the total flux of the injected model galaxy ($F_{\mathrm{\small model}}$) for over 350 simulated galaxies in the declination range -47\deg $\leq \delta \leq$ +8\deg. The data was cleaned to a residual flux threshold of 3.5 mJy in the minor CLEAN cycles.  The inverted yellow triangles represent the flux ratio of the model sources convolved with a perfect 30\arcsec\ Gaussian beam to that of the total flux of the injected source into the image cubes. \textbf{b):} Same as panel a), but now the sources were cleaned deeper to a residual flux threshold of 0.9 mJy. \textbf{c):} Shows the ratio of the integrated flux from the 12\arcsec\ and 30\arcsec\ model sources injected into to the image cubes and cleaned to a residual flux threshold of 3.5 mJy. \textbf{d):} Same as panel c), but now cleaned to a deeper residual flux threshold of 0.9 mJy. The points are color-coded based on the SNR of the 30\arcsec\ detections. The black solid line represents the expected one-to-one ratio, while the dashed red line shows the mean flux discrepancy of the distribution. }
    \label{fig:flux_compare_sims}
\end{figure*}

Panel \textbf{a)} in Figure~\ref{fig:flux_compare_sims} shows a plot of the ratio of the 30\arcsec\ integrated flux to the total flux of the injected model galaxy ($\frac{F_{30}}{F_{\textrm{model}}}$) against the injected model galaxy flux ($F_{\textrm{model}}$). The detections are color-coded based on their SNR values, which are computed as the total flux divided by the uncertainty in the flux measurement, both of which are provided by SoFiA in the source catalogue. As mentioned before, the CLEAN thresholds are set according to the default ASKAPSoft pipeline which is cleaning to a residual peak flux density threshold of 3.5 mJy.
We find that indeed the measured flux of the mock galaxies in the 30\arcsec\ resolution is consistently lower than the flux of the injected model galaxies. The mean flux discrepancy is $\sim 4$\%, while the discrepancy becomes more and more prominent towards the low-flux and low-SNR regime, where the offset can be as much as 15\%, corresponding closely to the discrepancy observed in the real data. 

Following this, we explored the observed flux discrepancy between the 12\arcsec\ and 30\arcsec\ data. Panel \textbf{c)} in Figure~\ref{fig:flux_compare_sims} shows the ratio of the 12\arcsec\ to 30\arcsec\ fluxes ($\frac{F_{12}}{F_{30}}$) plotted against the 30\arcsec\ fluxes of injected model galaxies. We find that the 12\arcsec\ fluxes are typically $\sim 10$\% higher compared to the 30\arcsec\ fluxes, which is comparable to what is observed in real ASKAP observations (about $\sim 15$\%). From these simulations, it is evident that the impact of the positive sidelobes on the uncleaned flux in the 12\arcsec\ resolution is quite significant. A similar observation has also been reported in \citet{Radcliffe23}, who undertook a simulation study to investigate the impact of asymmetric PSFs from interferometers on the recovered flux from sources. Their study points to the fact that non Gaussian dirty beams (PSFs) lead to consistent flux offsets compared to the flux of the injected model source. We note that our simulations are in agreement with their observations, wherein the flux is more discrepant for the marginally resolved, low SNR sources, while being somewhat more consistent for extended sources. 

As noted previously, this flux discrepancy may be an effect of incomplete cleaning, which results in the uncleaned flux being included into the final image cubes, impacting the final measured integrated flux. In order to investigate the impact of deeper cleaning, we cleaned the data to a peak residual flux threshold of $\sim 0.5\sigma = 0.9$ mJy in the minor cycles with an additional deep cleaning threshold set to 0.1 mJy. The deeper deconvolution thresholds lead to better flux recovery for both the 30\arcsec\ and 12\arcsec\ data sets, with the measured and injected fluxes in better agreement. This is shown in panels \textbf{b)} and \textbf{d)} in Figure~\ref{fig:flux_compare_sims}. 

To summarise, we note the following observations from the simulation study. 
\begin{itemize}
    \item We find that the 30\arcsec\ integrated flux is consistently lower by about 4\% compared to the integrated flux of the input model galaxy. However, at the low flux (or SNR) end, the flux discrepancy can be as high as 15\%, consistent with observations. This effect is observed in panel \textbf{a)} in Figure~\ref{fig:flux_compare_sims} (also Figure~\ref{fig:flux_compare_30arcsec}), were we observe that sources with an SNR $< 20$ show a higher flux discrepancy.
    \item We find that the 12\arcsec\ fluxes are on average consistently higher than the 30\arcsec\ fluxes by $\sim 4 - 10$\%, depending on a number of factors, including the SNR of the data, as well as how extended and bright the source is.
    \item The 12\arcsec\ and 30\arcsec\ fluxes for the Gaussian beam convolved data sets are consistent with each other for the high SNR (SNR $> 20$), while not surprisingly the 12\arcsec\ fluxes for the marginally resolved and/or low SNR sources is typically lower than the 30\arcsec\ fluxes.
    \item We note that by cleaning deeper in both the 30\arcsec\ and 12\arcsec\ data sets, we recover most of the flux, almost completely resolving the flux discrepancy. This suggests that deeper CLEANing thresholds are essential to fully recover the flux from WALLABY observations.
\end{itemize}

We do acknowledge that while care was taken to carry-out the simulations to reflect as closely as possible the ASKAPSoft pipeline, there are many subtle differences that might still impact the way the data is processed and hence the recovered fluxes. Such effects are likely to impact sources in the low-flux (-SNR) regime more than well resolved and higher flux sources. However, despite these caveats, the simulations do highlight the importance of deeper cleaning of the ASKAP observations in order to recover fluxes properly.

\subsection{Correcting the fluxes}
\label{subsec:flux_correction}

We showed in the previous section through simulations that in order to properly recover the total flux from a source, we need to clean much deeper (potentially 0.5$\sigma$ or deeper) than the current deconvolution thresholds set in ASKAPSoft, in addition to using source masks generated from shallower CLEAN runs for further cleaning. Another important point is the implementation of joint deconvolution routines that enable the visibility data from all ASKAP primary beams be jointly imaged and deconvolved, so that the data can be cleaned to the appropriate deeper CLEAN thresholds. However, such a system is still not in place in the current ASKAPSoft pipeline and furthermore will require significantly more computational resources to process the data. This means that one cannot fully clean the data and therefore the impact of the sidelobes on the residual flux will remain an issue for both the 12\arcsec\ and 30\arcsec\ data-sets. However, the WALLABY team is currently testing newly implemented parameters in ASKAPSoft's deconvolution algorithm and optimal CLEAN thresholds will be implemented for the full WALLABY survey accordingly.

In the interim, we statistically correct the 30\arcsec\ fluxes of Phase 2 sources by appropriately scaling their integrated fluxes to their corresponding single dish values from ALFALFA and HIPASS. We find that the data is best fit by a second order polynomial of the form

\begin{equation}
    \log_{10}\left( \frac{F_{\textrm{W}}}{F_{\textrm{SD}}} \right) = -0.006448 \phi^{2} + 0.103635 \phi - 0.439071
\end{equation}

where $\phi = \log_{10}\left( \frac{F_{\textrm{W}}}{\textrm{Jy Hz}} \right)$. The red line in the upper panel in Figure~\ref{fig:flux_compare_30arcsec} shows the fit to the data, while the bottom panel shows the corrected fluxes. We  see that this seems to systematically bring the flux level up, making it more consistent with the single-dish data. We note that we resorted to the new polynomial fit for the Phase 2 data, as the third order polynomial fit from PDR1 (dashed grey line in the upper panel in Figure~\ref{fig:flux_compare_30arcsec}) does not fit the data very well and seems to over-correct the fluxes in the low-flux end. The reason for the very different flux offsets observed between Phase 1 and Phase 2 data is likely stemming from the fact that Phase 2 observations utilised the holography-based primary beam correction as opposed to the use of a Gaussian primary beam correction in Phase 1, which will lead to systematic offsets in the flux. 

We note that these corrections have not been applied to the data products for each source included as part of this public data release, however, we have included the corrected fluxes in the catalogue and advise the users to be aware of this issue and apply the necessary correction to the fluxes when using the image cubes and moment maps for any analysis. The keywords \textbf{f\_sum\_corr} and \textbf{err\_f\_sum\_corr} in the source catalogue represent the corrected flux and error on the corrected flux, respectively. Similarly, the keyword \textbf{log\_m\_hi\_corr} represents the \h1 mass derived from the corrected flux and using the Hubble distance to the source. 

%%%%%%%%%%%%%%%%%%%%%%%%%%%%%%%%%%% Section: Kinematic modelling   %%%%%%%%%%%%%%%%%%%%%%%%%%%%%%%%%%%%%%%%%%%%%%%%%%%%%%%
\section{Kinematic Modelling}
\label{sec:3D_modelling}

One of the goals of WALLABY is to generate kinematic models for as many galaxies as possible.  For Phase 1 \citet{Deg2022} developed the WALLABY Kinematic Analysis Proto-Pipeline (\textsc{WKAPP}\footnote{\href{https://github.com/CIRADA-Tools/WKAPP}{WKAPP is available at https://github.com/CIRADA-Tools/WKAPP}}) that is optimized for the low resolution and signal-to-noise (S/N) of the standard 30\arcsec\ data.  It uses a combination of two different tilted ring (TR) modelling algorithms to generate reliable kinematic models from observed source cubelets. It was used to generate the 109 kinematic models of WALLABY Phase 1 and we use it here on both the 30\arcsec\ and 12\arcsec\ source cubelets.

Tilted-ring modelling treats a galaxy as a series of nested rings described by a number of observational parameters (center, systemic velocity, position angle, and inclination angle) and intrinsic ones (surface density, disk thickness, rotation velocity, and velocity dispersion).  This technique, introduced by \citep{rogstad74}, was first developed for 2D images and has been adapted to work with 3D data cubes.  There are a number of advantages to working in 3D including the ability to apply more complex models to well resolved, high S/N data (see for instance \citealt{Jorza2009,Khoperskov2014,DiTeoDoro2021,Jorzsa2021}). More relevant to the WALLABY context, 3D TR algorithms are also able to model galaxies at lower spatial resolutions across a wider range of disk geometries than equivalent 2D algorithms \citep[e.g.][]{Kamphuis2015, diTeodoro15, Lewis2019, Jones2021}.

While a full description of \textsc{WKAPP} is found in \citet{Deg2022} we will briefly describe the key points here.  \textsc{WKAPP} combines fits from two different 3D TR algorithms to generate its models -- Fully Automated \textsc{TiRiFiC} (FAT, \citealt{Kamphuis2015}), which itself is built on the Tilted Ring Fitting Code (TiRiFiC; \citealt{Jorza2007}); and the 3D-Based Analysis of Rotating Objects From Line Observations (\textsc{BBAROLO}; \citealt{diTeodoro15}).  Both codes are run in a `flat-disk' mode, where the observed geometry is constant across all rings. \citet{Deg2022} found that the differences between the two codes tended to be larger than the reported uncertainties of either algorithm.  As such, \textsc{WKAPP} uses half the difference between the models as the better estimate of the model uncertainty, which is applied to all galaxies with either a \textsc{SoFiA} \textit{ell\_maj} $\ge 2$ beams or an integrated $\log(S/N)\ge 1.25$.  The fits for each code are compared, and if both fits are reasonable, the two are averaged to create the final kinematic model.

In this section we describe the results of applying \textsc{WKAPP} to both the 30\arcsec\ and 12\arcsec\ data.  Section~\ref{subsec:NR_KinModels} focuses on the 30\arcsec\ Phase 2 data and how the models compare to Phase 1.  Section~\ref{subsec:HR_KinModels} focuses on the 12\arcsec\ data and how those models compare to the 30\arcsec\ models.

\subsection{Normal Resolution Modelling}
\label{subsec:NR_KinModels}

While Phase 2 contains many more detections than Phase 1 ($\sim1800$ unique detections compared to the $\sim600$ unique detections of Phase 1), these galaxies tend to be further away and smaller in size than the Phase 1 observations.  Of these Phase 2 galaxies, only 275 have the requisite size and S/N to attempt kinematic modelling.  For comparison, Phase 1 contains 209 unique galaxies that satisfy the \textit{ell\_maj} $\ge 2$ or $\log(S/N)\ge1.25$ criteria.  Table \ref{tab:KinModelTable} lists the sources, the galaxies that satisfied the modelling attempt criteria, and the total number of kinematic models for Phase 2.  Considering only those galaxies where kinematic modelling is attempted, the 30\arcsec\ Phase 2 sources have a success rate of 45\%, which is comparable to Phase 1.

\begin{table}[]
    \centering
    \begin{tabular}{|c|c|c|c|}
        \hline 
        Source Release & $N$ sources & $N$ attempted & $N$ models \\
        \hline 
        NGC 4808 TR1 & 231 & 38 & 18\\ 
        NGC 5044 TR1 & 353 & 55 & 23\\
        NGC 5044 TR2 & 630 & 76 & 36\\
        NGC 5044 TR3 & 1326 & 185 & 85 \\
        Vela TR1  & 203 & 43 & 17\\
        Total 30\arcsec\ & 2743 & 397 & 179 \\
        Unique 30\arcsec\ & 1827 & 275 & 127 \\
        \hline 
        NGC 4808 High-Res TR1 & 12 & 12 & 5 \\
        NGC 5044 High-Res TR1 & 55 & 54 & 21 \\
        Vela High-Res TR1 & 13 & 13 & 4 \\
        Total 12\arcsec\ & 80 & 79 & 30 \\
        \hline 
    \end{tabular}
    \caption{The number of sources, attempts, and successful models in each release (where TR refers to Team Release). Note that there are no double sources in the 12\arcsec\ data so a `Unique' 12\arcsec\ row is the same as the `Total' 12\arcsec\ row.}
    \label{tab:KinModelTable}
\end{table}

\begin{figure*}
    %\centering
    \vspace*{-1.5cm}
    \includegraphics[width=\textwidth]{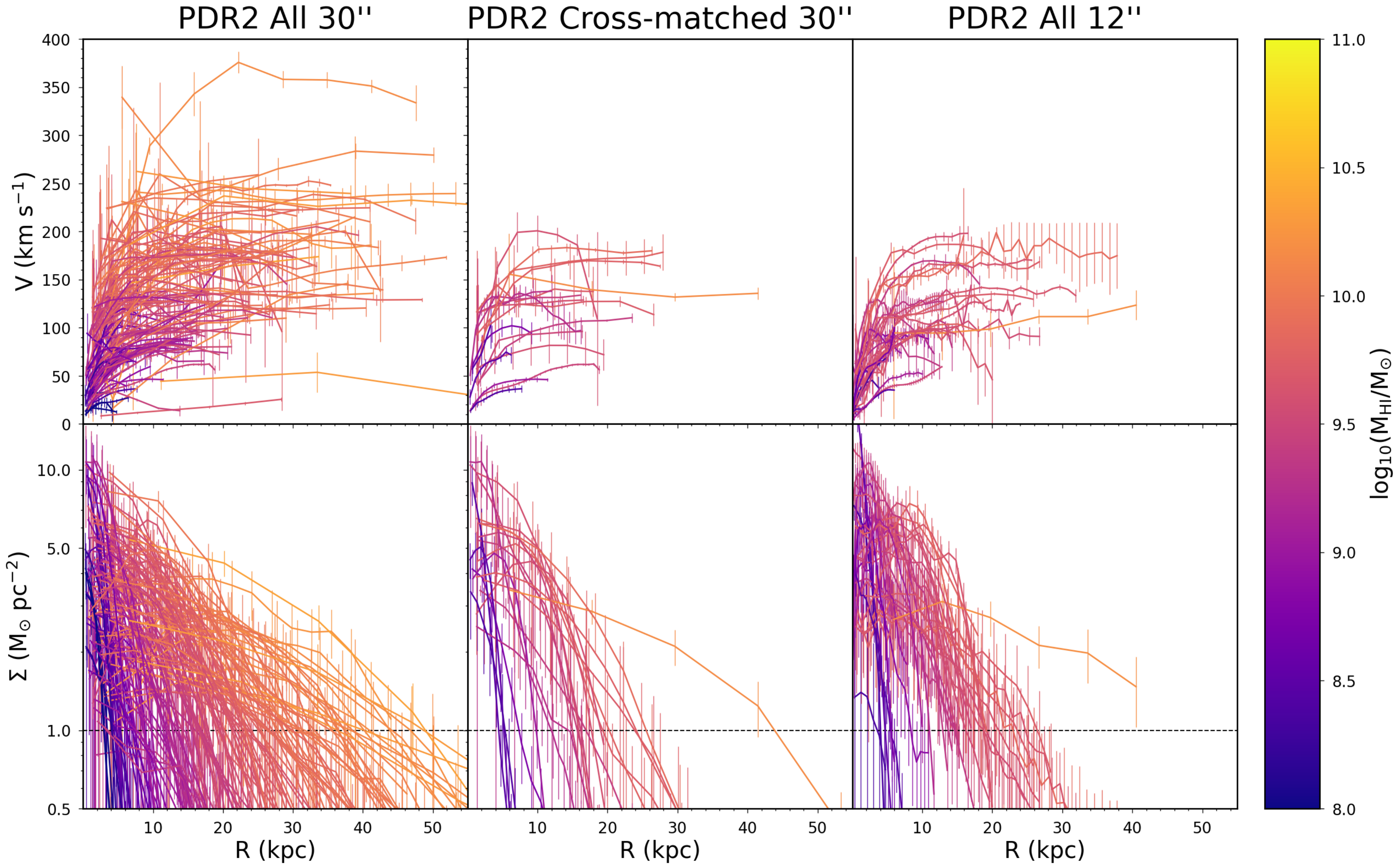}
    \caption{The rotation curves (top row) and deprojected surface density profiles (bottom row) for Phase 2.  The left hand panels shows the models for all 30\arcsec\ data while the right hand panels show the models for the 12\arcsec\ data.  The middle column shows the 30\arcsec\ models for galaxies that also have a model from their 12\arcsec\ data. The dashed horizontal line in the surface density panels is at 1 M$_{\odot}$ pc$^{-2}$, which is the standard value used to define $R_{\textrm{\h1}}$. }
    \label{fig:RCs_And_SDs}
\end{figure*}

Using \textsc{WKAPP}, the final catalogue of Phase 2 kinematic models contains 127 unique galaxies.  The left-hand panels of Figure \ref{fig:RCs_And_SDs} shows the kinematic models for these 127 models.  These models span a wide range of rotation velocities and extents, including a number of low-mass dwarfs.  It is worth noting that the surface densities are calculated through ellipse-fitting on the moment 0 map using the averaged model geometry.  As such the deprojected profiles shown in Figure \ref{fig:RCs_And_SDs} have not been corrected for beam smearing effects, which may be important for some applications.

\subsection{High Resolution Kinematic Modelling}
\label{subsec:HR_KinModels}

\textsc{WKAPP} was developed with an eye towards the 30\arcsec\ WALLABY data, and it is not clear that this approach is appropriate for the higher resolution 12\arcsec\ data.  Nonetheless, we have applied the proto-pipeline data to the high resolution data and obtained 27 kinematic models from the 80 detections.  The rotation curves and deprojected surface density profiles for these models are shown in the right-hand panels of Figure \ref{fig:RCs_And_SDs}.

Comparing the left and right panels of Figure \ref{fig:RCs_And_SDs}, it is clear that the distribution of 12\arcsec\ and 30\arcsec\ models are not the same. The 12\arcsec\ models are biased towards higher rotation velocities and \h1 masses compared to the 30\arcsec\ models (which show a wider range of velocities and \h1 masses). This is evidenced by the fact that very few modelled galaxies in the 12\arcsec\ resolution have rotation velocities lower than 80 \kms\ in the outer/flat parts of their rotation curves. This is likely due to HIPASS preferentially finding relatively nearby gas-rich galaxies, which are typically observed to have higher rotation velocities.

The middle panel of Figure \ref{fig:RCs_And_SDs} shows the 30\arcsec\ models for galaxies that also have a 12\arcsec\ model.  Some galaxies in the 12\arcsec\ sample that are successfully modelled do not have an equivalent 30\arcsec\ model.  Thus, while there are 27 12\arcsec\ models, there are only 18 cross-matched 30\arcsec\ models.  Comparing the middle and right columns of Figure \ref{fig:RCs_And_SDs} reveals that the rotation curves are broadly equivalent for the cross-matched models.  However the 12\arcsec\ models tend to be truncated relative to the 30\arcsec\ data.

\begin{figure*}
    \vspace*{-3.2cm}
    \centering
    \includegraphics[width=\textwidth]{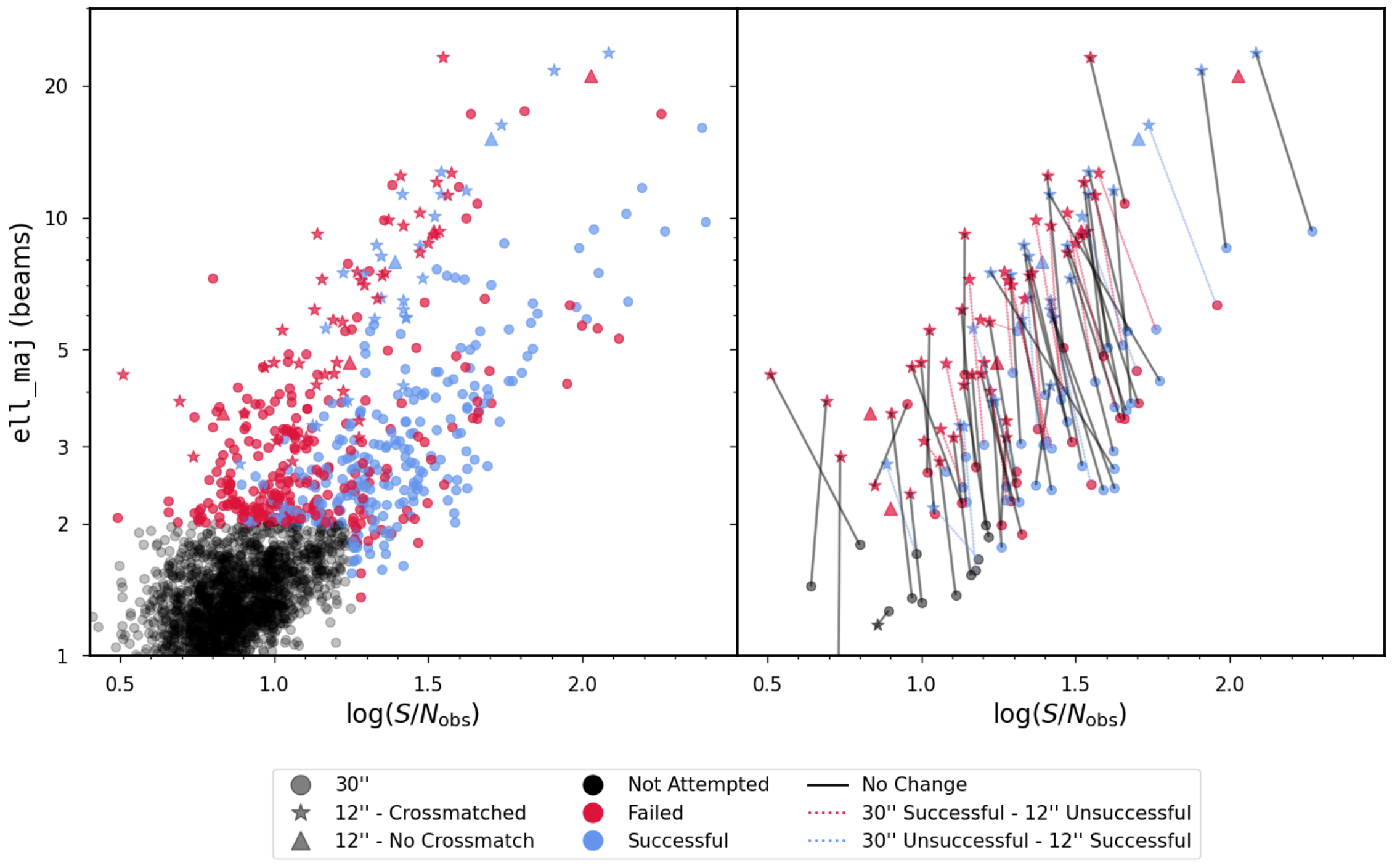}
    \caption{The size and integrated S/N of the Phase 2 sources.  The circles show the 30\arcsec\ detections, while the stars and triangles shows the 12\arcsec\ detections.  The different 12\arcsec\ symbols indicate whether there is a cross-matched 30\arcsec\ source for the 12\arcsec\ source (stars) or not (triangles).  The black, red, and blue points indicate galaxies where kinematic modelling was not attempted, attempted and failed, or successfully modelled respectively.  The left hand panel shows all Phase 2 detections, while the right hand panel only shows the 12\arcsec\ sources and their crossmatched 30\arcsec\ counterpart (if a crossmatched source exists).  In the right hand panel the lines connect the cross-matched sources.  Occasionally a 30\arcsec\ source is broken into two different sources and will have two lines originate from the source.  If the kinematic modelling result has not changed (failed for both or successful for both), the line is black.  If the 30\arcsec\ source is kinematically modelled while the 12\arcsec\ source is not the line is red, and when the situation is reversed the line is blue.}
    \label{fig:KinModels_SizeSN}
\end{figure*}

To gain a better understanding of the Phase 2 models and the 12\arcsec\ data, Figure \ref{fig:KinModels_SizeSN} shows the \textsc{SoFiA} \textit{ell\_maj} parameter and integrated S/N for the Phase 2 sources.  As noted in \citet{Deg2022}, there is a clear relationship between the size and the integrated S/N.  In the left-hand panel that shows all Phase 2 sources, it is clear that there is a diagonal limit above which no successful models are generated.  For a given size, a higher S/N leads to a higher chance of kinematically modelling a galaxy.  Conversely, larger galaxies with the same S/N as smaller galaxies are more difficult to model with \textsc{WKAPP}.

Focusing on the 12\arcsec\ detections and the cross-matched 30\arcsec\ sources reveals a number of interesting behaviours.  Firstly, the majority of the cross-matched sources have approximately the same $\log(S/N)$.  Secondly, the approximate size of the modelled disc has not increased by a factor of 2.5. This is expected as the smaller beam size results in a worse column density sensitivity, which means that the most extended gas will be below the noise limit. This, combined with the fact that the beam smearning effects are minimised in the 12\arcsec\ resolution also explains the decreased radial extent of the 12\arcsec\ kinematic models seen in Figure \ref{fig:RCs_And_SDs}.

A third, and perhaps more important result is apparent in Figure \ref{fig:KinModels_SizeSN}.  Only 8 galaxies do not have a 30\arcsec\ kinematic model and a successful 12\arcsec\ model (indicated by blue lines in Figure \ref{fig:KinModels_SizeSN}).  By contrast there are 18 sources that were successfully modelled using their 30\arcsec\ data that were not modelled with the 12\arcsec\ data.  These results show that, for \textsc{WKAPP}, the increased noise of the 12\arcsec\ data leads to poorer results in terms of kinematic modelling despite the increased resolution.  It is important to note here that \textsc{WKAPP} is being run in precisely the same way for the 12\arcsec\ as for the 30\arcsec\ data.  If a more tailored approach were adopted it is possible that the kinematic modelling would be significantly more successful.  Additionally, the increased resolution brings many of the galaxies into the regime where various 2D algorithms are applicable.  Collapsing the data to moment maps effectively increases the S/N and may lead to greater success than the 3D approach of \textsc{WKAPP}.  These ideas will be explored in a future work.

\section{Data access}
\label{sec:data_access}

The WALLABY Pilot Survey Phase 2 data and associated catalogues are available to the public through the CSIRO ASKAP Science Data Archive (CASDA) and the Canadian Astronomy Data Centre (CADC). The data release is similar to Public Data Release 1 and includes all the 30\arcsec\ source data products, kinematic models and respective catalogues. In addition, in this release we are also including the high-resolution 12\arcsec\ data products, kinematic models and catalogue. We also provide descriptions and details on the various data products, data quality issues and list details of the various column names in the catalogues. 

We note that the source catalogue in this release also includes all detections from the Public Data Release 1 from \citet{westmeier2022} for easy  accessibility to both DR1 and DR2 detections. Furthermore, the new catalogue will include all the relevant updated columns such as the corrected fluxes (\textbf{f\_sum\_corr} and \textbf{err\_f\_sum\_corr}) as well as the corrected \h1 masses (\textbf{log\_mi\_hi\_corr}) for both the DR1 and DR2 samples making it convenient for the user to use the corrected values.

The combined footprint A and B mosaics are available on CASDA via \textbf{\url{https://doi.org/10.25919/hg66-4v60}}. These are very large (typically $\sim$600 GB) and we recommend that users interact with these via the CASDA cutout service. These cutouts can be made either through the CASDA Data Access Portal (DAP) or by using the Simple Image Access Protocol (SIAP) coupled with the Serverside Operations for Data Access (SODA) protocol. In the second case, the user interacts using a Python script or Jupyter notebook to select the region and channel range of interest. Additionally, the CASDA module of the Astropy Astroquery package\footnote{\href{https://astroquery.readthedocs.io/en/latest/casda/casda.html}{https://astroquery.readthedocs.io/en/latest/casda/casda.html}} can also generate cutouts.

Users can access the 30\arcsec\ data via CASDA using the following links for the various types of data-sets. a) Source data products (including moment maps, cubelet, channel map, source mask and spectra) and complete source catalogue:~\textbf{\url{https://doi.org/10.25919/qw7w-tn96}}; b) 30\arcsec\ kinematic modelling data products and catalogue:~\textbf{\url{https://doi.org/10.25919/7w8n-9h19}}.

The 12\arcsec\ source data products, which includes all SoFiA source data products (moment maps, cubelet, channel map, source mask and spectra), kinematics models and catalogue (including kinematic modelling parameter values) can be accessed via CASDA using the following link:~\textbf{\url{https://doi.org/10.25919/47tr-k441}}.

All the above data products for both the 30\arcsec\ and 12\arcsec\ data can be accessed via CADC through a TAP service using ADQL queries. For more details on how to access the data through CADC we refer the reader to the Public Data Release 1 papers (\citealt{westmeier2022};~\citealt{Deg2022}). Furthermore, users can also get detailed instructions and links to the data releases through WALLABY's data access page\footnote{\url{https://wallaby-survey.org/data/}}.

%%%%%%%%%%%%%%%%%%%%%%%%%%%%%%%%%%%% Section: Summary and Future   %%%%%%%%%%%%%%%%%%%%%%%%%%%%%%%%%%%%%%%%%%%%%%%%%%%%%%%%%%%%%%
\section{Summary and Future}
\label{sec:Summary_Future}

In this data release paper, we present the catalogue, data products including moment maps and spectra for over 1800 galaxies from the WALLABY Pilot Survey Phase 2. The observations were carried out on three selected fields which include the NGC 5044, NGC 4808 and Vela groups. The total observed sky area is $\sim 180$ deg$^2$ and the redshift limit corresponding to $z \sim 0.09$. The median rms noise levels in the data cubes is $\sim 1.7$ mJy, which is close to the expected theoretical noise for the WALLABY observations. This translates to a $5\sigma$ column density sensitivity of $\sim  9.1\times10^{19}(1 + z)^4$ cm$^{-2}$ assuming a 30\arcsec\ beam and a 20 \kms channel width. 

In addition to the default 30\arcsec\ data products, in Phase 2 we have also presented the high-resolution 12\arcsec\ cut-outs of select HIPASS galaxies demonstrating the true potential of WALLABY to produce high spatial and spectral resolution \h1 observations of several thousand galaxies (including all HIPASS galaxies) in the 5-year survey period, thereby forming the largest sample of high spatial resolution \h1 maps of galaxies until the SKA-mid begins observations. As such, these high-resolution cut-outs carry immense legacy value. 

We highlighted the significant improvement in the quality of the data compared to Phase 1 which is mainly attributed to the fact that the Pilot Phase 2 fields were selected in a way as to avoid bright continuum sources, but also due to the introduction of the holography-based primary beam correction for the ASKAP observations, which results in more accurate fluxes for the sources. It is to be noted that there is ongoing work to implement appropriate ``peeling" techniques into the ASKAPSoft data reduction pipeline in order to properly subtract residual continuum that is associated with bright continuum sources, likely improving the quality of the data significantly. 
While the data quality in general is very good, we note the observed flux discrepancy in the ASKAP observations. The issue was first highlighted in the Phase 1 paper \citep{westmeier2022}, wherein the integrated flux of the 30\arcsec\ WALLABY detections were observed to be $\sim 15$\% lower than the corresponding single-dish flux. This was alluded to improper deconvolution and the impact of residual sidelobes still present in the image cubes. In order to fully understand this issue, we undertook simulations of ASKAP observations and injected model galaxies by varying their properties such as flux and size and find that up on performing the source finding using SoFiA, the simulated galaxies in the 30\arcsec\ resolution are indeed observed to show consistently lower flux compared to the flux of the injected model galaxies. We attribute this to the contribution of the uncleaned flux in the data, which is impacted by the severe negative sidelobes that systematically brings down the integrated flux. We also note that marginally-resolved and/or low-SNR sources are more severely impacted by this. 

Furthermore, we also note that the integrated flux of the 12\arcsec\ sources is observed to be consistently higher than their 30\arcsec\ counterparts, which again is attributed to the impact of uncleaned flux in the data and which has the imprint of the highly non-Gaussian 12\arcsec\ ASKAP dirty beam with strong positive sidelobes. This uncleaned flux therefore artificially boosts the flux of the 12\arcsec\ detections to about $\sim$ 15\% depending on a number of factors including the SNR and spatial extent of the source. In order to minimise the impact of the uncleaned flux on the data, going forward for the full WALLABY survey, it is necessary to set appropriate cleaning thresholds and making sure that the thresholds are reached during the clean cycles. In addition, a two stage cleaning approach involving a shallow clean followed by a deeper cleaning using a source mask might result in better flux recovery. This will of course considerably increase the time and resources required to process the data, however, such a scheme may be implemented in the ASKAPSoft spectral-line imaging pipeline given the ASKAP observations are now being processed in the new upgraded Pawsey HPC \textit{Setonix}, which is capable of handling large data volumes. 

\begin{acknowledgement}
We would like to thank the anonymous referee for their useful comments which improved the clarity of this paper. We would also like to sincerely thank Minh Huynh (CSIRO) for all the efforts towards releasing the data onto CASDA.

This scientific work uses data obtained from Inyarrimanha Ilgari Bundara / the Murchison Radio-astronomy Observatory. We acknowledge the Wajarri Yamaji People as the Traditional Owners and native title holders of the Observatory site. CSIRO’s ASKAP radio telescope is part of the Australia Telescope National Facility (https://ror.org/05qajvd42). Operation of ASKAP is funded by the Australian Government with support from the National Collaborative Research Infrastructure Strategy. ASKAP uses the resources of the Pawsey Supercomputing Research Centre. Establishment of ASKAP, Inyarrimanha Ilgari Bundara, the CSIRO Murchison Radio-astronomy Observatory and the Pawsey Supercomputing Research Centre are initiatives of the Australian Government, with support from the Government of Western Australia and the Science and Industry Endowment Fund.

This research used the facilities of the Canadian Astronomy Data Centre operated by the National Research Council of Canada with the support of the Canadian Space Agency.

The Canadian Initiative for Radio Astronomy Data Analysis (CIRADA) is funded by a grant from the Canada Foundation for Innovation 2017 Innovation Fund (Project 35999) and by the Provinces of Ontario, British Columbia, Alberta, Manitoba and Quebec, in collaboration with the National Research Council of Canada, the US National Radio Astronomy Observatory and Australia’s Commonwealth Scientific and Industrial Research Organisation.

This paper includes archived data obtained through the CSIRO ASKAP Science Data Archive, CASDA (http://data.csiro.au).

WALLABY acknowledges technical support from the Australian SKA Regional Centre (AusSRC) and Astronomy Data And Computing Services (ADACS).

This research has made use of the NASA/IPAC Extragalactic Database (NED), which is funded by the National Aeronautics and Space Administration and operated by the California Institute of Technology.

Parts of this research were supported by the Australian Research Council Centre of Excellence for All Sky Astrophysics in 3 Dimensions (ASTRO 3D), through project number CE170100013.

KS acknowledges funding from the Natural Sciences and Enginneeing Research Council of Canada.

LC acknowledges support from the Australian Research Council via the Discovery Project funding scheme (DP210100337)
 
NY acknowledges the fellowship of the China Postdoctoral Science Foundation (grant: 2022M723175, GZB20230766).

PK is partially supported by the BMBF project 05A23PC1 for D-MeerKAT. 

LVM acknowledges financial support from the grant CEX2021-001131-S funded by MCIU/AEI/ 10.13039/501100011033, from the grant PID2021-123930OB-C21 funded by MCIU/AEI/ 10.13039/501100011033 and by ERDF/EU

\end{acknowledgement}

\bibliography{references}

\appendix

\section{ASKAPSoft imaging parameters for the 30\arcsec\ and 12\arcsec\ data reduction}
\label{appendix:imaging_params}
In Table~\ref{tab:imaging_params} we list some relevant ASKAPSoft imaging parameters used to process both the default 30\arcsec\ and 12\arcsec\ data. For more details on the definition of each of the parameters we refer the reader to the ASKAPSoft User Documentation.\footnote{\url{https://www.atnf.csiro.au/computing/software/askapsoft/sdp/docs/current/index.html}}

\begin{table*}
\caption{Important ASKAPSoft imaging, pre-conditioning, deconvolution and tapering parameters for the 30\arcsec\ and 12\arcsec\ data processing}
\begin{tabular}{llll}
\hline\hline
Parameter & 30\arcsec\ & 12\arcsec\ & Description \\
\hline
SPECTRAL\_IMAGE\_MAXUV & 2000 & 7000 & Maximum UV distance (in metres) \\
 & & & to apply in the data selection step \\
SPECTRAL\_IMAGE\_MINUV & 12   & 12   & Minimum UV distance (in metres)  \\ 
 &  &  & to apply in the data selection step \\
NUM\_PIXELS\_SPECTRAL    & 1536 & 384  & No. of spatial pixels along the side for \\
 & & & the image cubes \\
CELLSIZE\_SPECTRAL      & 6    & 2    & The spatial pixel size for the image cubes \\
CLEAN\_SPECTRAL\_SCALES & [0, 20, 60, 120] & [0, 10, 30, 60] & Set of scales (in pixels) to use with the \\
 & & & multi-scale clean \\
CLEAN\_SPECTRAL\_PSFWIDTH & 3172 & 256 & The width of the psf patch used \\
 & & & in the minor cycle (in pixels) \\
CLEAN\_SPECTRAL\_THRESHOLD\_MINORCYCLE & [45\%, 3.5mJy, 0.5mJy] & [45\%, 3.5mJy, 0.5mJy] & Threshold for the minor cycle loop \\
CLEAN\_SPECTRAL\_THRESHOLD\_MAJORCYCLE & 0.5mJy & 0.5mJy & The target peak residual. Major cycles \\
& & &  stop if this is reached \\
PRECONDITIONER\_LIST\_SPECTRAL & [Wiener, GaussianTaper] & [Wiener, GaussianTaper]  & List of preconditioners to apply \\ 
PRECONDITIONER\_SPECTRAL\_WIENER\_ROBUSTNESS &  0.5  &  0.5 &  Robustness value for the Wiener filter \\
PRECONDITIONER\_SPECTRAL\_GAUSS\_TAPER &  [30arcsec, 30arcsec, 0deg] & [12arcsec, 12arcsec, 0deg] & Size of the Gaussian taper FWHM (in arcsec) \\
& & & and position angle in degrees \\
\hline
\end{tabular}
\label{tab:imaging_params}
\end{table*}

\section{SoFiA parameters for the 30\arcsec\ and 12\arcsec\ source finding runs}
\label{appendix:SoFiA_params}

In Table~\ref{tab:SoFiA_params_30and12}, we list some important SoFiA parameter values used for the source finding runs for the 30\arcsec\ and 12\arcsec\ data sets.

\begin{table}
\caption{SoFiA parameter values for the 30\arcsec\ and 12\arcsec\ source finding runs.}
\begin{tabular}{lcc}
\hline\hline
Parameter & 30\arcsec\ & 12\arcsec\ \\
\hline
{\tt scaleNoise.windowZ} & 51 & 51 \\
{\tt scaleNoise.windowXY} & 51 & 75 \\
{\tt scfind.kernelsXY} & 0, 5, 10 & 0, 6, 12 \\
{\tt scfind.kernelsZ} & 0, 3 ,7, 15, 31 &  0, 3, 7, 15, 31 \\
{\tt scfind.threshold} & 3.8 & 4.0 \\
{\tt scfind.replacement} & 2.0 & 1.4 \\
{\tt linker.radiusXY} & 2 & 2 \\
{\tt linker.radiusZ} & 3 & 3 \\
{\tt linker.minSizeXY} & 5 & 6 \\
{\tt linker.minSizeZ} &  5 & 5 \\
{\tt reliability.threshold} & 0.7 & 0.7 \\
{\tt reliability.scaleKernel} & 0.2 & 0.2 \\
{\tt reliability.minSNR} & 3.0 & 3.0 \\
\hline
\end{tabular}
\label{tab:SoFiA_params_30and12}
\end{table}

\section{Manual Inspection Workflow}
\label{appendix:manual_inspection_workflow}

A source finding pipeline run generates detections and associated data products, which are then added to a database. %(\S\ref{appendix:source_find_pipeline_database}) in the {\tt Detection} and {\tt Product} tables, with reference to a {\tt Run} and various {\tt Instance} entries for a particular run of the pipeline. When a detection is marked as a WALLABY source, an entry is created in the {\tt Source} table with a link to the detection via a {\tt SourceDetection} entry. 
The database is populated through a manual process whereby all detections are visually examined by the WALLABY team to ensure that artefacts, false detections, and duplicates are removed. A web portal has been developed for conveniently executing the various stages of this inspection workflow.

\begin{figure}
    %\centering
    \hspace*{-0.65cm}
    \includegraphics[width=1.3\columnwidth]{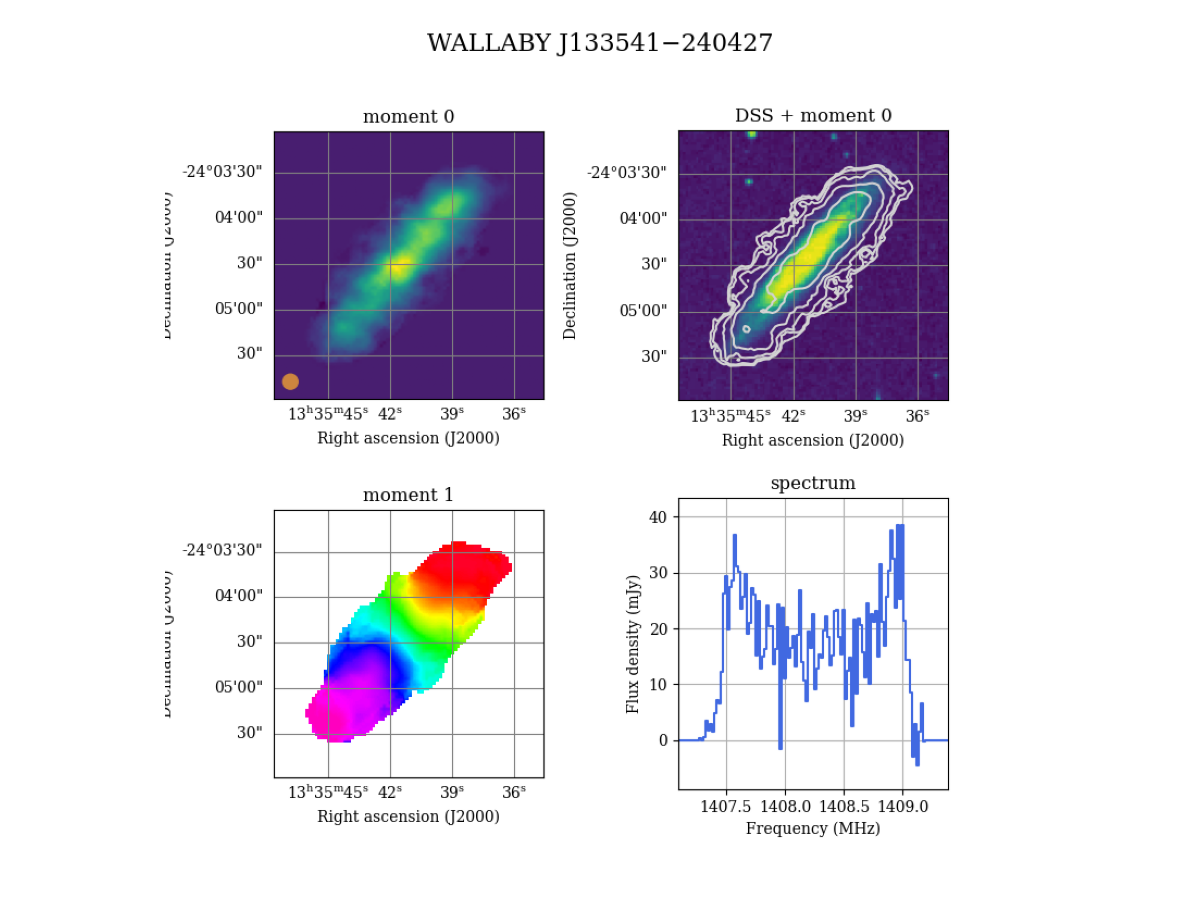}     
    \caption{Summary figure presenting the moment 0, moment 1 map, spectra, and optical DSS image of a source. These summary figures, along with properties of the detection from the source finding application are used by the WALLABY team to identify and remove false detections.}
    \label{fig:sampledetection}
\end{figure}

To determine whether a detection is a real source, a WALLABY team member is presented with key detection properties (such as flux, RA, Dec) in a table, and a visual summary of data products (moment 0 and 1 maps, spectra, and an overlay of the moment 0 \h1 contours on to a Digitized Sky Survey (DSS) optical image. An example of the summary figure is shown in Figure~\ref{fig:sampledetection}. Detections that pass this first check are selected as potential genuine sources. %A pre-release entry in the {\tt Source} table, and {\tt SourceDetection} link between this detection and the pre-release source, are created. Only detections with pre-release {\tt Source} entry and {\tt SourceDetection} entries are considered in subsequent stages of this workflow.

In the case where there is a staggered approach to selecting sources, for example, overlapping regions of the sky subsequently processed by the source finding pipeline may give rise to duplicate detections of already accepted sources. This is the case for the NGC~5044 field, where overlapping regions are shown in Figure~\ref{fig:source_finding_strategy} in the darker shade of green in the corners of the central $4^{\circ} \times 4^{\circ}$ processing regions. In such cases, an external cross matching routine is performed by the inspection workflow allowing the WALLABY team to identify and handle potential duplicate detections. The spatial and spectral locations of the detections from each new source finding run are compared against accepted source entries in the catalogue. If they are within a tight spatial or spectral threshold ($\Delta_{\rm spat} \pm 5$\arcsec, $\Delta_{\rm spec} \pm 0.05$ MHz), they are automatically marked as duplicates and are removed from the database. If the candidate is within a lenient spatial and spectral threshold ($\Delta_{\rm spat} \pm 90$\arcsec, $\Delta_{\rm spec} \pm 2$ MHz), they are marked for an additional visual inspection step, before being accepted as a genuine detection and assigned a WALLABY source name. Once this workflow is completed, the accepted sources are ready for release.

\section{Output source catalogue}
\label{appendix:output_catalogue}

Table~\ref{tab:detectiontable} provides details of all the parameters that are included in the source catalogue for all PDR2 detections. Two additional parameters are listed in the 12\arcsec\ source catalogue that represent the 12\arcsec\ integrated flux corrected to the original 30\arcsec\ integrated flux and the associated statistical uncertainty.

\begin{table*}
\caption{List of parameters in the source catalogue.}
\begin{tabular}{lllll}
\hline\hline
Name & Type & Unified Content Descriptor & Units & Description \\
\hline
name & char & meta.id; meta.main & - & WALLABY source name (WALLABY Jhhmmss+/-ddmmss) \\
ra & double & pos.eq.ra; meta.main & deg & Right ascension (J2000) of centroid position \\
dec & double & pos.eq.dec; meta.main & deg & Declination (J2000) of centroid position \\
freq & double & em.freq; meta.main & Hz & Barycentric frequency of centroid position \\
f\_sum & double & phot.flux; meta.main & Jy*Hz & Integrated flux within 3D source mask \\
err\_f\_sum & double & stat.error; phot.flux & Jy*Hz & Statistical uncertainty of integrated flux \\
f\_sum\_corr & double & phot.flux; meta.main & y*Hz & Integrated flux within 3D source mask corrected to the single-dish integrated flux \\
err\_f\_sum\_corr & double & stat.error; phot.flux & Jy*Hz & Statistical uncertainty of corrected integrated flux \\
rms & double & instr.det.noise & Jy/beam & Local RMS noise level near source \\
w20 & double & spect.line.width; meta.main & Hz & Spectral line width at 20\% of the peak (w20) \\
w50 & double & spect.line.width & Hz & Spectral line width at 50\% of the peak (w50) \\
kin\_pa & double & pos.posAng & deg & Position angle of kinematic major axis \\
rel & double & stat.probability & - & Statistical reliability of detection from 0 to 1 \\
qflag & double & meta.code.qual & - & Source finding quality flag \\
kflag & double & meta.code & - & Kinematic model flag \\
n\_pix & double & meta.number; instr.pixel & - & Number of pixels in 3D source mask \\
f\_min & double & phot.flux.density; stat.min & Jy/beam & Lowest flux density value within 3D source mask \\
f\_max & double & phot.flux.density; stat.max & Jy/beam & Highest flux density value within 3D source mask \\
ell\_maj & double & phys.angSize & pix & Major axis size of ellipse fitted to moment 0 map \\
ell\_min & double & phys.angSize & pix & Minor axis size of ellipse fitted to moment 0 map \\
ell\_pa & double & pos.posAng & deg & Position angle of ellipse fitted to moment 0 map \\
ell3s\_maj & double & phys.angSize & pix & Same as ell maj but > 3 sigma pixels only and equal weight \\
ell3s\_min & double & phys.angSize & pix & Same as ell min but > 3 sigma pixels only and equal weight \\
ell3s\_pa & double & pos.posAng & deg & Same as ell pa but > 3 sigma pixels only and equal weight \\
x & double & pos.cartesian.x & pix & Centroid position in x \\
err\_x & double & stat.error;pos.cartesian.x & pix & Statistical uncertainty of centroid position in x \\
y & double & pos.cartesian.y & pix & Centroid position in y \\
err\_y & double & stat.error;pos.cartesian.y & pix & Statistical uncertainty of centroid position in y \\
z & double & pos.cartesian.z & pix & Centroid position in z \\
err\_z & double & stat.error;pos.cartesian.z & pix & Statistical uncertainty of centroid position in z \\
x\_min & double & pos.cartesian.x; stat.min & pix & Lower end of bounding box in x \\
x\_max & double & pos.cartesian.x; stat.max & pix & Upper end of bounding box in x \\
y\_min & double & pos.cartesian.y; stat.min & pix & Lower end of bounding box in y \\
y\_max & double & pos.cartesian.y; stat.max & pix & Upper end of bounding box in y \\
z\_min & double & pos.cartesian.z; stat.min & pix & Lower end of bounding box in z \\
z\_max & double & pos.cartesian.z; stat.max & pix & Upper end of bounding box in z \\
comments & char & meta.note & - & Comments on individual sources \\
team\_release & char & meta.dataset; meta.main & - & Internal WALLABY team release identifier \\
dist\_h & double & pos.distance & Mpc & Local Hubble distance derived from the barycentric source frequency \\
log\_m\_hi & double & phys.mass & $M_{\odot}$ & \h1 mass $\log(M_{\rm HI}/M_{\odot})$ derived from the local Hubble distance \\
log\_m\_hi\_corr & double & phys.mass & $M_{\odot}$ & \h1 mass $\log(M_{\rm HI}/M_{\odot})$ derived using the single-dish corrected integrated flux \\
\hline
f\_sum\_corr\_30 & double & phot.flux; meta.main & Jy*Hz & Integrated 12\arcsec\ flux within 3D source mask corrected to the 30\arcsec\ integrated flux \\
err\_f\_sum\_corr\_30 & double & stat.error; phot.flux & Jy*Hz & Statistical uncertainty of 12\arcsec\ integrated flux corrected to the 30\arcsec\ integrated flux \\
\hline
\end{tabular}
\label{tab:detectiontable}
\end{table*}

\section{Dirty beams (PSFs) from simulations}
\label{appendix2}

\begin{figure*}[h!]
    \centering
    \includegraphics[width=1\columnwidth]{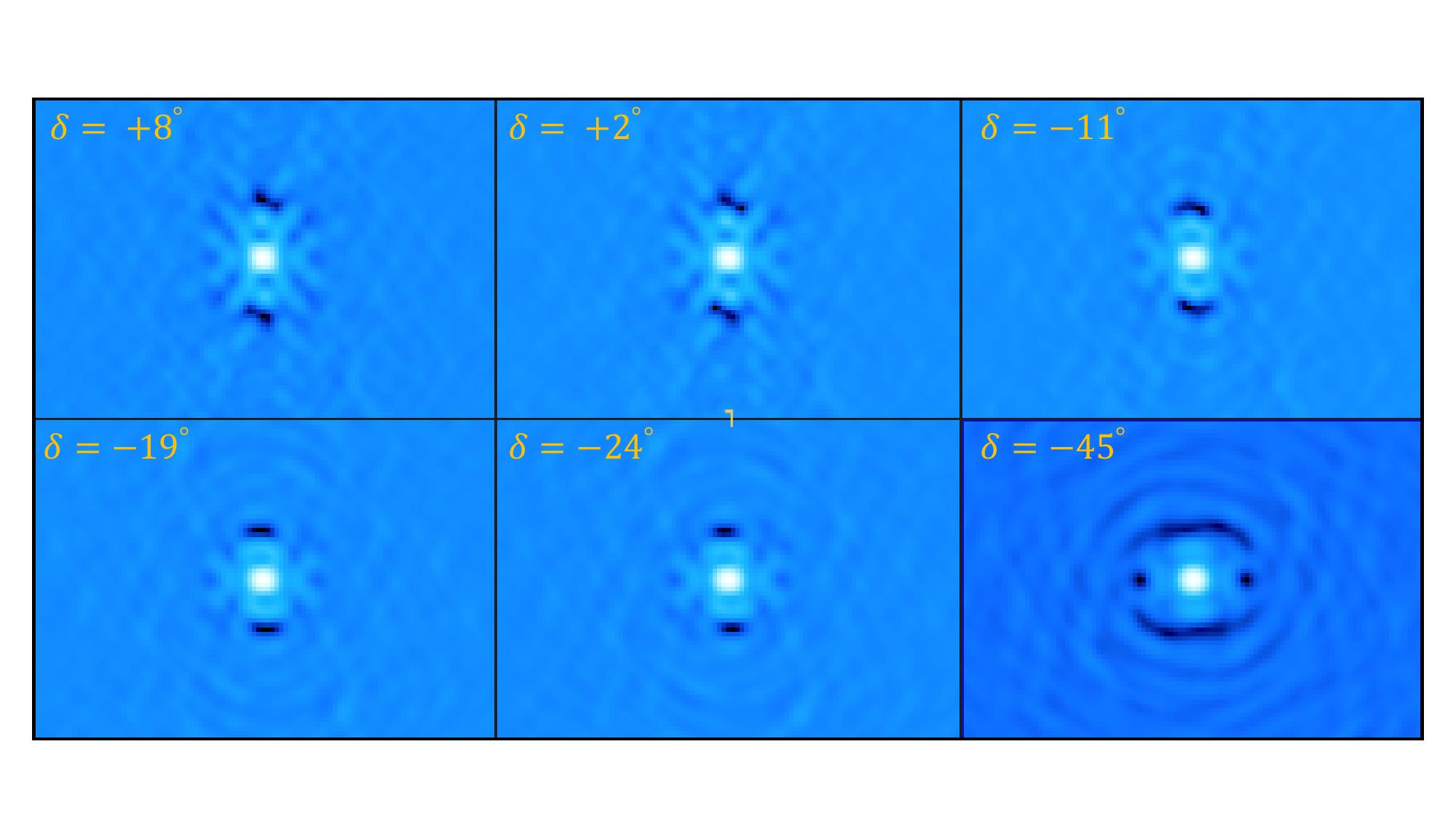}     
    \caption{The 12\arcsec\ dirty beams for various declinations from the simulations. }
    \label{fig:12arc_PSF_sims}
\end{figure*}

\begin{figure*}[h!]
    \centering
    \includegraphics[width=1\columnwidth]{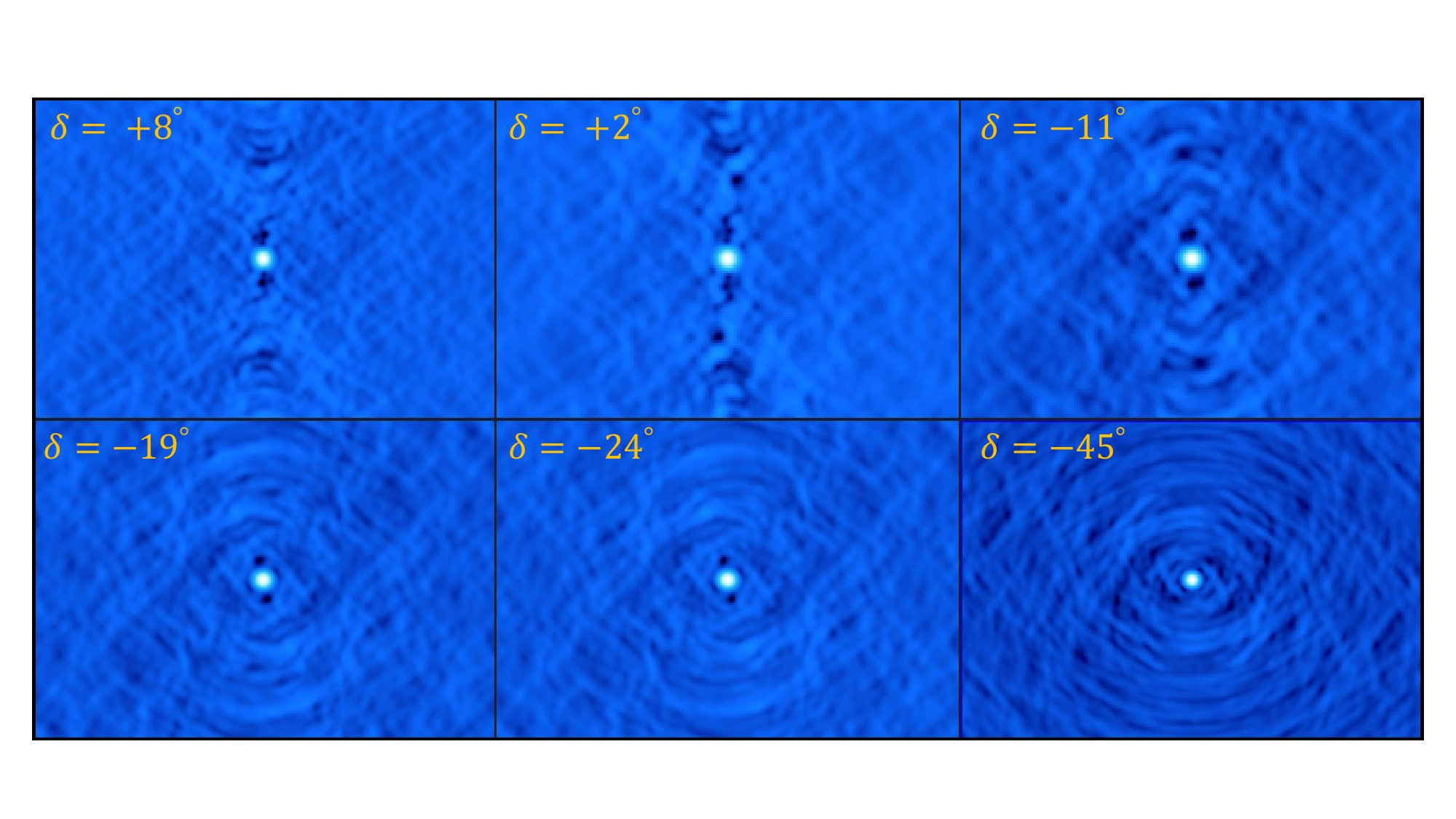}     
    \caption{The 30\arcsec\ dirty beams for various declinations from the simulations. }
    \label{fig:30arc_PSF_sims}
\end{figure*}

In Figure.~\ref{fig:12arc_PSF_sims} we show the simulated PSFs of the 12\arcsec\ beam for various declinations using MIRIAD tasks. We used a robust parameter of 0 and applied appropriate tapering in order to generate a dirty beam that is approximately 12\arcsec. The positive sidelobes associated with the 12\arcsec\ dirty beam is very significant, making the beam highly non-Gaussian-like. In the case of the 30\arcsec\ PSFs (see Figure~\ref{fig:30arc_PSF_sims}), while the central part of the beam is more Gaussian-like, there are significant negative sidelobes associated with the dirty beams.

%%%% Have moved the tables to the appendices!! %%%%%%%%%%%%%%%
\end{document}